\documentclass{jfm}
\usepackage{graphicx}
\usepackage{epstopdf, epsfig}

\shorttitle{Reopening modes of an elasto-rigid channel}
\shortauthor{L. Duclou\'e, A. L. Hazel, A. B. Thompson and A. Juel}

\title{Reopening modes of a collapsed elasto-rigid channel}

\author{Lucie Duclou\'e\aff{1}
  \corresp{\email{lucie.ducloue@manchester.ac.uk}},
  Andrew L. Hazel\aff{2}, 
  Alice B. Thompson\aff{2}
 \and Anne Juel\aff{1}}

\affiliation{\aff{1}Manchester Centre for Nonlinear Dynamics and School of Physics and Astronomy - University of Manchester, Oxford Road, Manchester M13 9PL, UK
\aff{2}Manchester Centre for Nonlinear Dynamics and School of Mathematics - University of Manchester, Oxford Road, Manchester M13 9PL, UK}

\begin{document}

\maketitle

\begin{abstract}
Motivated by the reopening mechanics of strongly collapsed airways, we study the steady propagation of an air finger through a collapsed oil-filled channel with a single compliant wall. In a previous study using fully-compliant elastic tubes, a `pointed' air finger was found to propagate at high speed and low pressure, which, if clinically accessible, offers the potential for rapid reopening of highly collapsed airways with minimal tissue damage~\citep{heap2008anomalous}. The mechanism underlying the selection of that pointed finger, however, remained unexplained.\\
In this paper, we identify the required selection mechanism by conducting an experimental study in a simpler geometry: a rigid rectangular Hele-Shaw channel with an elastic top boundary. The constitutive behaviour of this elasto-rigid channel is non-linear and broadly similar to that of an elastic tube, but unlike the tube the channel's cross-section adopts self-similar shapes from the undeformed state to the point of first near wall contact. The ensuing simplification of the vessel geometry enables the systematic investigation of the reopening dynamics in terms of increasing initial collapse.\\
We find that for low levels of initial collapse, a single centred symmetric finger propagates in the channel and its shape is set by the tip curvature. As the level of collapse increases, the channel cross-section develops a central region of near opposite wall contact, and the finger shape evolves
smoothly towards a `flat-tipped' finger whose geometry is set by the strong depth gradient near the channel walls. We show that the flat-tipped mode of reopening is analogous to the pointed finger observed in tubes. Its propagation is sustained by the vessel's extreme cross-sectional profile at high collapse, while vessel compliance is necessary to stabilise it. A simple scaling argument based on the dissipated power reveals that this reopening mode is preferred at higher propagation speeds when it becomes favourable to displace the elastic channel wall rather than the viscous fluid.\\
\end{abstract}


\section{Introduction}
\label{intro}
The dynamics of viscous, free-surface flows in the presence of compliant boundaries is controlled by a three-way competition between capillary, elastic and viscous forces. In this paper, our focus is the role of this competition in pulmonary biomechanics and, in particular, during the first breath of a newborn infant, when air must penetrate a branched network of extensively collapsed, liquid-filled compliant airways~\citep{bland2001loss} in order to reach the alveoli and establish gas exchange. The penetration is thought to occur via the propagation of a long air finger~\citep{hodson1991first} that displaces the occluding liquid (working against viscous resistance) and also reopens the compliant airways (working against elastic forces), as observed by~\citet{macklem1970stability} during the inflation of a collapsed, excised feline bronchiole. \\

In fact, two-way elastocapillary interactions are sufficient to deform thin elastic structures in contact with a droplet or thin film of fluid, provided that the stresses exceed the Young's modulus of the structure~\citep{roman2010elasto}. The Young's modulus is a fixed material constant, but capillary stresses scale inversely with the radius of curvature of the interface. Thus, the relative increase in capillary forces at the microscale means that these elastocapillary deformations can be used for applications in microfabrication~\citep{pokroy2009self} and soft robotics~\citep{borno2006transpiration}. Elastocapillarity can cause the collapse of elastic tubes during capillary rise~\citep{hoberg2014elastocapillary} and, in the lungs, it is associated with closure of the smallest airways, which occurs regularly as part of the breathing cycle~\citep{burger1968airway}: the compliant liquid-lined airways buckle under a fluid-elastic instability, and become obstructed by a plug of liquid~\citep{halpern1992fluid, heil1999minimal}. Extensive or permanent pulmonary collapse may occur under pathological conditions affecting either the properties of the lining fluid (e. g. asthma or pulmonary oedema) or the airways themselves (e. g. emphysema). In contrast to the first-breath scenario, only a limited region of the lung is collapsed. Nonetheless, in either scenario the airways must be (re)opened quickly with minimal damage to the surrounding tissue. Hence, understanding the mechanics of airway reopening is essential in order to design effective ventilation techniques.\\

\begin{figure}
\begin{center}
\includegraphics[scale=0.59]{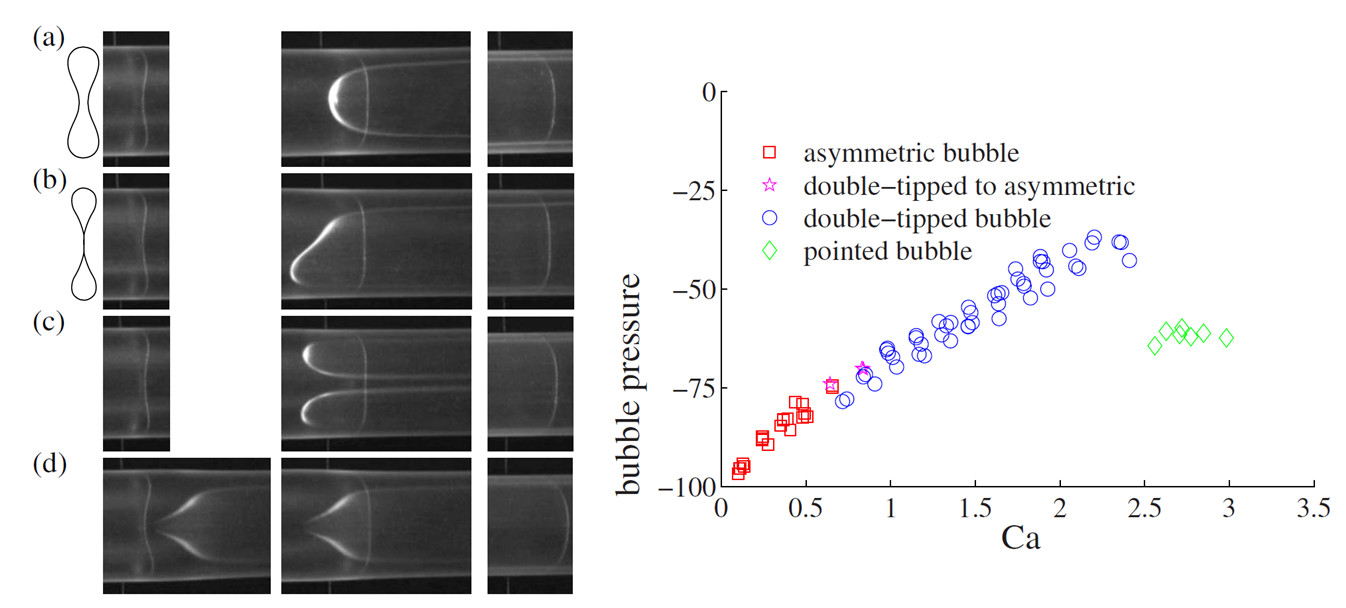}
\end{center}
\caption{(\textit{left panel}) Typical images of the finger shapes observed during the reopening of a collapsed elastic tube: (\textit{a}) round-tipped symmetric finger; (\textit{b}) asymmetric finger; (\textit{c}) double-tipped finger; (\textit{d}) pointed finger. The finger speed increases from (\textit{a}) to (\textit{d}). The left hand side pictures and schematic drawings show the tube's initial (static) cross section, which is the same from (\textit{b}) to (\textit{d}). The level of initial collapse is low in (\textit{a}), and high enough for near opposite wall contact in (\textit{b}) to (\textit{d}). (\textit{right panel}) Air finger pressure (scaled by the capillary pressure over the tube radius) as a function of the dimensionless finger speed (capillary number $\mathrm{Ca}=\mu U/\sigma$, with $\mu$ the fluid's dynamic viscosity, $U$ the finger speed and $\sigma$ the fluid's surface tension) for a given level of high initial tube collapse. \textit{(figures reproduced from~\citet{heap2008anomalous} with permission)}.\label{fig:pointed}}
\end{figure}

An elastic vessel's collapse under external load can be modelled according to a constitutive relationship often referred to as a ``tube law'', which relates the pressure difference $P_{in}-P_{out}$ across the vessel wall to a measure of the vessel collapse, e.g. the normalised cross-sectional area. Upon decreasing transmural pressure, the cross-sectional area of the vessel typically decreases monotonically, but non-linearly. A tube of initially circular cross-section~\citep{flaherty1972post, shapiro1977steady} first buckles into an elliptical shape followed by a succession of two-lobed configurations (illustrated in figure~\ref{fig:pointed} (\textit{a})) until the opposite boundaries of the tube come into contact at a critical transmural pressure, and the cross-section adopts a dumbbell shape (as seen in figure~\ref{fig:pointed} (\textit{b})). Beyond this critical pressure, the effective stiffness of the tube (the gradient of the tube law) changes dramatically because small reductions in cross-sectional area require large increases in transmural pressure. \\

The reopening of a liquid-filled, collapsed airway relies on an increase in pressure at the inner wall of the airway, which depends on the injected air pressure working against viscous and capillary stresses. The resulting flow of air displacing the viscous fluid is predominantly controlled by the capillary number $\mathrm{Ca}= \mu U/\sigma$, which is the ratio of viscous to surface tension stresses. Here, $\mu$ is the viscosity of the fluid, and $\sigma$ its surface tension. The capillary number is the main control parameter as long as the geometry is sufficiently small so that gravitational and inertial effects can be neglected.  \\

The first benchtop experimental study of airway reopening was performed by~\citet{gaver1990effects}: a liquid-filled, thin-walled polyethylene tube of negligible bending stiffness was collapsed into a ribbon shape, held under longitudinal tension to introduce resistance to deformation, and reopened by injecting air at a constant pressure. Reopening occurred through the propagation of a steadily advancing air finger, which peeled apart the walls of the tube. The propagation of this finger was captured by two-dimensional modelling and numerical computations~\citep{jensen2002steady, grotberg2004biofluid}, using a linear constitutive relation between the pressure and the reopening height of the tube. The linear constitutive relation did not account for the observed stiffness changes in the tube law of real elastic tubes; in particular, a change in the level of initial collapse was equivalent to a simple rescaling in the transverse direction. The influence of the nonlinear tube law was examined in three-dimensional numerical simulations~\citep{hazel2003three} performed under imposed symmetry about the vertical and horizontal mid planes of the tube. The general behaviour of the three-dimensional system was the same as the two-dimensional model, but the nonlinear tube law introduced a dependence on the initial level of collapse of the tube into the relationship between finger speed and air pressure. More strongly collapsed tubes were found to be ``easier to reopen'' in the sense that lower pressures were required to support steadily propagating states because smaller volumes of fluid needed to be displaced. Restrictions in the numerical method meant that very strongly collapsed tubes could not be investigated. In all the above experiments, models and simulations, the tube was reopened by a single family of steadily propagating, symmetric, round-tipped fingers.\\

Alternative modes of finger propagation, illustrated in figure~\ref{fig:pointed}, were uncovered in subsequent reopening experiments of fluid-filled elastic tubes in the limit of strong collapse~\citep{juel2007reopening, heap2008anomalous, heap2009bubble}. When the tube was collapsed into a two-lobe configuration, the symmetric round-tipped finger (\textit{a}) was replaced with `asymmetric' fingers (\textit{b}) as the finger speed increased. For tubes collapsed beyond opposite wall contact, a very thin layer of wetting liquid remained between opposite boundaries, unlike the solid-solid contact in a dry tube. In this regime of strong collapse, `asymmetric', `double-tipped' (\textit{c}) and `pointed' fingers (\textit{d}) were successively obtained for increasing finger speed~\citep{heap2008anomalous}. Moreover, the symmetric finger was not observed. Asymmetric and double-tipped fingers exhibited a similar linear increase of the finger pressure with $\mathrm{Ca}$, while the transition to a pointed finger was associated with a discontinuous reduction in finger pressure, as shown in figure~\ref{fig:pointed}. Hence, the pointed finger is of potential clinical interest, because it enables fast reopening while its low pressure minimises tissue damage. The transitions from symmetric to asymmetric and double-tipped fingers were understood to occur in order to minimise the viscous resistance at the finger tip~\citep{heap2009bubble}. By contrast, the mechanism by which the pointed finger reopened the tube, infiltrating the thin liquid layer in the middle region of near-opposite wall contact, remained unexplained. \\

The propagation of the pointed finger in the thin, central region of the tube where viscous resistance is maximum, suggests that the inflation of the tube is essential to sustain this finger~\citep{heap2009bubble}. In general, the reopening tube interacts with the free-surface displacement flow to select the finger shape via two geometric features: (i) non-uniform viscous resistance set by the non-uniform cross-sectional depth of the tube, and (ii) axial taper generated by the inflation of the tube in the reopening region. The constitutive behaviour of the tube prevented a systematic analysis of the coupling between these features because of the multiplicity of shapes adopted by the tube cross-section with increasing collapse. Hence, subsequent studies have investigated decoupled problems in simplified geometries by considering either: (i) the propagation of an air finger in a fully rigid channel with a cross-section of non-uniform depth, or (ii) front propagation in a cell of uniform initial thickness with a flexible upper boundary that inflates in response to air injection. We briefly summarize the results of these studies below.\\

The injection of air into a liquid-filled rigid-walled Hele-Shaw channel of rectangular cross-section gives rise to the classical viscous fingering instability, which results in the propagation of a single, centred symmetric finger~\citep{saffman1958penetration, tabeling1987experimental}. The width of the finger is a function of the modified capillary number $\alpha^2\mathrm{Ca}$, where $\alpha=W/H$ is the aspect ratio of the channel's cross-section, with $W$ and $H$ being the cross-sectional width and depth, respectively. Several different modes of finger propagation, including stable asymmetric and oscillatory modes, were observed in channels with an axially uniform, centred step-like constriction on the bottom boundary~\citep{de2009tube, franco2016sensitivity}, a geometry that approximates the cross-section of a statically collapsed elastic tube. In addition, a depth-averaged model~\citep{thompson2014multiple, franco2016sensitivity} revealed a range of unstable solutions, but none adopted a pointed finger shape. By investigating taller constrictions that reflect strongly collapsed elastic cross-sections, we shall show in section~\ref{sec:highColl} that
pointed finger shapes can be obtained numerically in the depth-averaged model of rigid constricted channels, but that those numerical solutions are unstable, supporting the conjecture that local inflation and the consequent taper is necessary to sustain stable pointed fingers.\\

The influence of wall compliance on viscous fingering has been investigated in a radial Hele-Shaw cell by~\citet{pihler2012suppression}, who showed that the onset of viscous fingering can be delayed considerably when the upper bounding plate of the Hele-Shaw cell is replaced by an elastic membrane. Wall elasticity not only affects the onset of the instability but also has a strong impact on the structure of the fingers that subsequently develop. The entire interface propagates and the instability develops as a ring of short constant-depth fingers. The inflation of the membrane stabilises the interface by slowing down the advancing finger front and creating a locally convergent fluid-filled gap in which the interface propagates. Numerical simulations have shown that such a tapered cell geometry gives rise to viscous and capillary contributions to the stabilisation of the interface~\citep{pihler2013modelling}. In rigid-walled rectangular Hele-Shaw channels, a converging depth gradient has also been shown to enhance the range of capillary numbers for which a flat interface remains linearly stable~\citep{al2012control, al2013controlling}. However, steadily propagating fingers of finite amplitude have not been observed in rigid tapered channels.\\

In this paper, we investigate the reopening of a fluid-filled collapsed vessel in terms of the coupling between a highly collapsed initial cross-section and the axial taper due to the inflation of the vessel. We focus on a novel, rectangular elasto-rigid Hele-Shaw channel, in which the upper boundary is replaced by an elastic membrane. We find that this setup has the two advantages compared to the collapsible tube: (i) the ``tube law'' is qualitatively similar to that of the tube, but the shapes adopted by the channel cross-section under decreasing transmural pressure are self-similar (see section~\ref{sec:exp}); and (ii) the flow dynamics are expected to be quasi two-dimensional and can be approximated by a depth-averaged model. Our experimental set-up is presented in section~\ref{sec:exp}, in which we also characterise the constitutive behaviour of the elasto-rigid channel. In section~\ref{sec:collapse}, we characterise the reopening of the elasto-rigid channel at fixed capillary number as a function of the initial level of collapse, starting from the rectangular cross-section at zero transmural pressure. We find that the finger shape broadens as the level of initial collapse is increased, and smoothly transitions towards a reopening mode, the `flat-tipped' finger, which is analogous to the pointed finger in the tube for high levels of initial collapse. We examine the propagation of the flat-tipped finger as a function of $\mathrm{Ca}$ in section~\ref{sec:Ca}, where we identify a transition to an alternative mode that is associated with a discontinuous jump in pressure. We rationalise this transition with a scaling argument which provides a good estimate of the critical capillary number for the transition, both in our channel and in the tube geometry used by~\citet{heap2008anomalous}.\\

\section{Experimental set-up and methods}
\label{sec:exp}

\subsection{Apparatus and experimental procedure}
\begin{figure}
\begin{center}
\includegraphics[height=6.5cm]{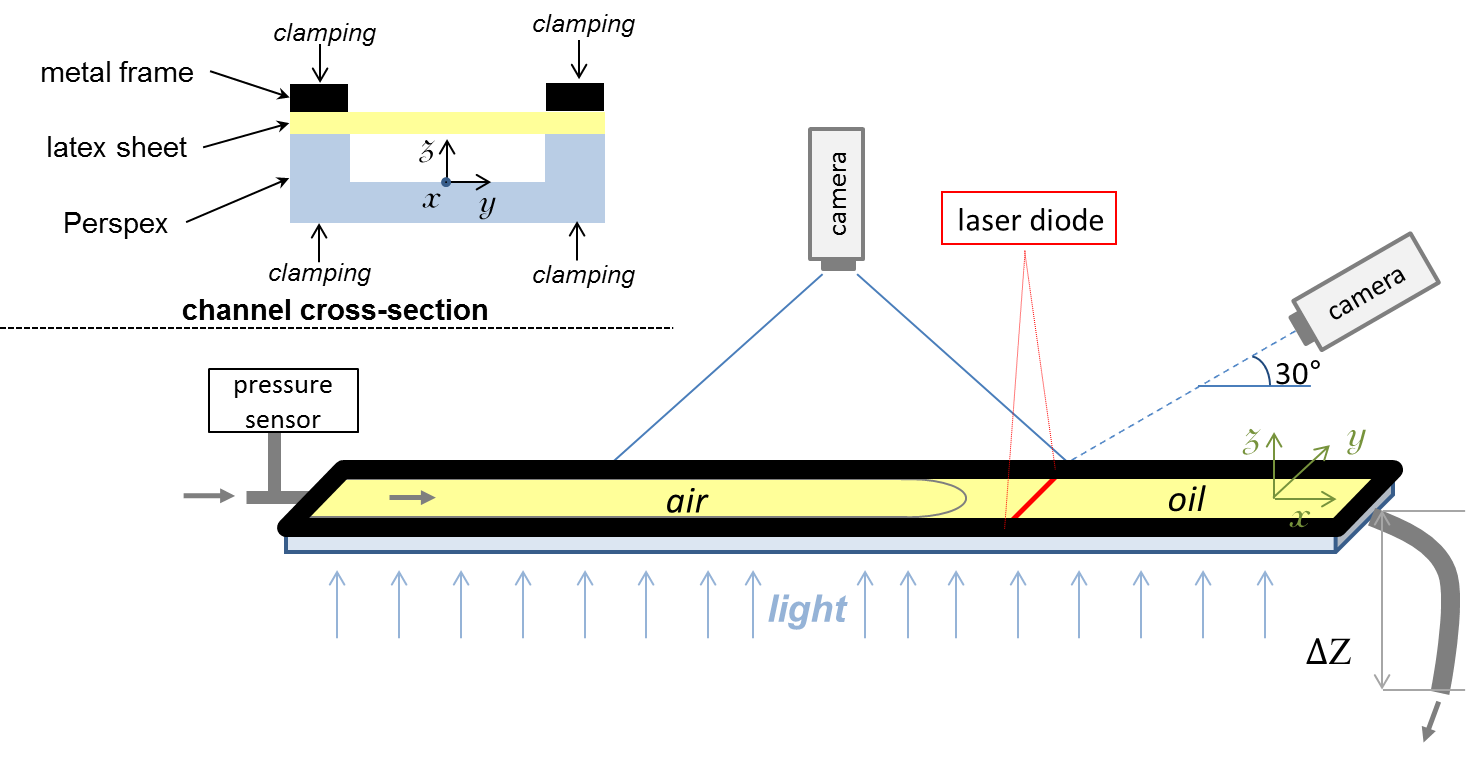}
\caption{Schematic diagram of the experimental set-up; the inset in the top left corner shows the channel cross-section. A latex sheet is clamped on top of a rigid rectangular thin-gapped channel. Two cameras allow for visualisation of the finger shape and speed, and membrane vertical displacement. A differential pressure sensor records the pressure of the propagating air finger. The height difference $\Delta Z$ between the channel and the exit of the outlet tubing sets the hydrostatic pressure in the oil prior to air injection.\label{fig:setup}}
\end{center}
\end{figure}

The working part of the experimental set-up is a thin-gapped rectangular channel with a flexible top boundary shown schematically in figure~\ref{fig:setup}. A rigid rectangular channel of length $L=60$~cm, width $W=30.00\pm 0.02$~mm and depth $H=1.00\pm 0.01$~mm was milled in a 2.5~cm thick block of transparent Perspex. The roughness of the channel floor resulting from milling is smaller than 10~$\mathrm{\mu m}$. The top wall of the channel consisted of a sheet of translucent latex (Supatex) of uniform thickness $d = 0.34\pm 0.01$~mm, Young's modulus $E = 1.44 \pm 0.05$~MPa and Poisson's ratio $\nu = 0.5$~\citep{pihler2012suppression}. The latex sheet was cut to larger dimensions than the channel (63~cm $\times$ 10~cm) and clamped on the four edges of the channel with a rigid 1.2 cm thick rectangular metal frame. The frame was screwed to the Perspex block on one long edge and the two short ones, and the fourth edge was clamped using 11 evenly spaced G-clamps. When setting up the channel, the sheet was first laid flat onto the Perspex base, before the metal frame was positioned on top of it and screwed to the Perspex over the long edge to secure the sheet. Tension was applied along the width of the sheet by uniformly hanging weight (with a total mass of 868 g) to the free long edge. The short edges of the frame were then screwed in and the G-clamps placed on the unscrewed side to finish holding the sheet edges, before the hanging weight was removed. This resulted in the sheet lying flat on top of the empty channel (as illustrated in the inset of figure~\ref{fig:setup}), and being pre-stressed with a tension of approximately 40 $\mathrm{kPa}$, assuming that no significant pre-stress variations were generated when clamping the sheet. This pre-stress remains small compared to the Young's modulus of the latex. The channel was levelled to better than $0.1^\circ$.\\
 
The propagation of the air finger was recorded via a CCD camera (Dalsa Genie, of resolution 1400$\times$1024 pixels, fitted with a 12~mm $f$/1.8 lens, acquiring images at a rate of 4 to 8 frames per second depending on the finger speed) mounted on a tripod half a metre above the channel, which imaged a 40~cm long section of the channel located 10~cm downstream of the channel inlet. A custom made closed light box consisting of opalescent Perspex panels evenly diffusing the light of an array of LED diodes provided bright and uniform back-lighting. The vertical displacement of the elastic sheet was monitored by shining a laser sheet vertically onto the latex across the channel width 36~cm downstream of the inlet. The position of this bright line was recorded by a second CCD camera (Pixelink, $1280\times1024$ pixels, fitted with a 35~mm $f$/2 lens and capturing images at a rate of 5 to 22 fps depending on the finger speed) placed 10~cm above the channel and oriented at 30.0$^\circ \pm 0.1^\circ$ from the horizontal. The position of the imaging regions was chosen to ensure that a steady propagation mode had been reached. \\

The channel was filled with silicone oil (Basildon Chemicals Ltd.) of viscosity $\mu = 0.099$ Pa.s, surface tension $\sigma = 21$ $\mathrm{mN/m}$ and density $\rho = 973$ $\mathrm{kg/m^3}$ at the laboratory temperature of $21\pm 1^\circ$C. The oil fully wetted both the Perspex and the latex sheet. Prior to each experiment, the channel was collapsed by imposing a uniform pressure loading onto the sheet. The pressure difference $P_{in}-P_{out}$ between the inside of the channel and the atmosphere (transmural pressure $\Delta P$) was chosen by setting the static pressure in the oil: after injecting oil to fill the channel, the channel inlet was closed and the outlet connected to a length of tubing open to the atmosphere. The height difference $\Delta Z$ (see figure~\ref{fig:setup}) between the open end of the tubing and the channel resulted in a hydrostatic pressure $P_{in}=P_{out}-\rho g \Delta Z$ (where $g$ is the acceleration due to gravity) in the oil inside the channel, which set the transmural pressure. We then waited for the oil volume in the channel to adjust. Images of the channel cross-sectional profiles were taken regularly during that process in order to monitor when the final state was reached, which took between 45~min and 6~h depending on the transmural pressure. We checked for the smallest pressure differences, which took the longest to adjust, that the shape was not modified by waiting for much longer periods (up to 24~h).\\

Once the equilibrium position of the membrane was reached, an experimental run was performed by injecting air at a constant flow rate via the channel inlet. The air flow was supplied by a syringe pump (KDS 200) fitted with gas-tight syringes (Hamilton) for flow rates between 10 and 150~mL/min. For higher flow rates (between 100 and 300~mL/min), the syringe pump could no longer be used and gas was supplied by a compressed nitrogen cylinder whose flow rate was adjusted by a fine needle valve and monitored by a mass air flow meter (Red-Y Smart Meter PCU1000, Voegtlin Instruments AG). The flow rate from the compressed gas source could be replicated to within 3\% using this procedure. A differential pressure sensor (Honeywell, $\pm$~5"~$\mathrm{H_2O}$) recorded the pressure $P_b$ of the propagating air finger. Accurate measurements of this pressure required the removal of any oil menisci in the gas line connecting the channel to the sensor. This was achieved by injecting a small volume of air (approximately 1~mL) in the channel just before the start of the experiment, and ensuring continuous connection of this precursor bubble to the air supply and pressure sensor by trapping the small amount of oil left in the gas supply line in a short section of enlarged tubing upstream of the channel inlet. \\

Images from the camera placed above the channel were processed using Matlab R2013a routines which subtracted the background from the original images, optimized the contrast of the resulting images to use the whole grey scale range and ran a Canny edge-detection algorithm (built-in \texttt{edge} function) to extract the outline of the propagating finger. The ratio of the distance travelled by the tip to the time interval between consecutive frames was computed and averaged over the whole visualisation window to calculate the finger tip speed $U$. The finger width $\lambda$ (scaled by the channel width) was measured on each frame over a length of approximately 2.5 cm located at a fixed distance (1.5 cm) behind the finger tip, and averaged over the whole visualisation window. The images of the laser line recorded by the side camera during finger propagation were analysed by detecting the position of the line relative to the fixed edges of the channel. A sub-pixel resolution of 20~$\mathrm{\mu m}$ on the vertical position of the line was achieved by fitting the grey scale intensity of each pixel column across the line to a Gaussian profile to locate the point of maximum intensity, as described by~\citet{pihler2015displacement}.\\

\subsection{Constitutive behaviour of the elasto-rigid channel}
\label{sec:constit}
\begin{center}
\begin{figure}
\includegraphics[height=4.75cm]{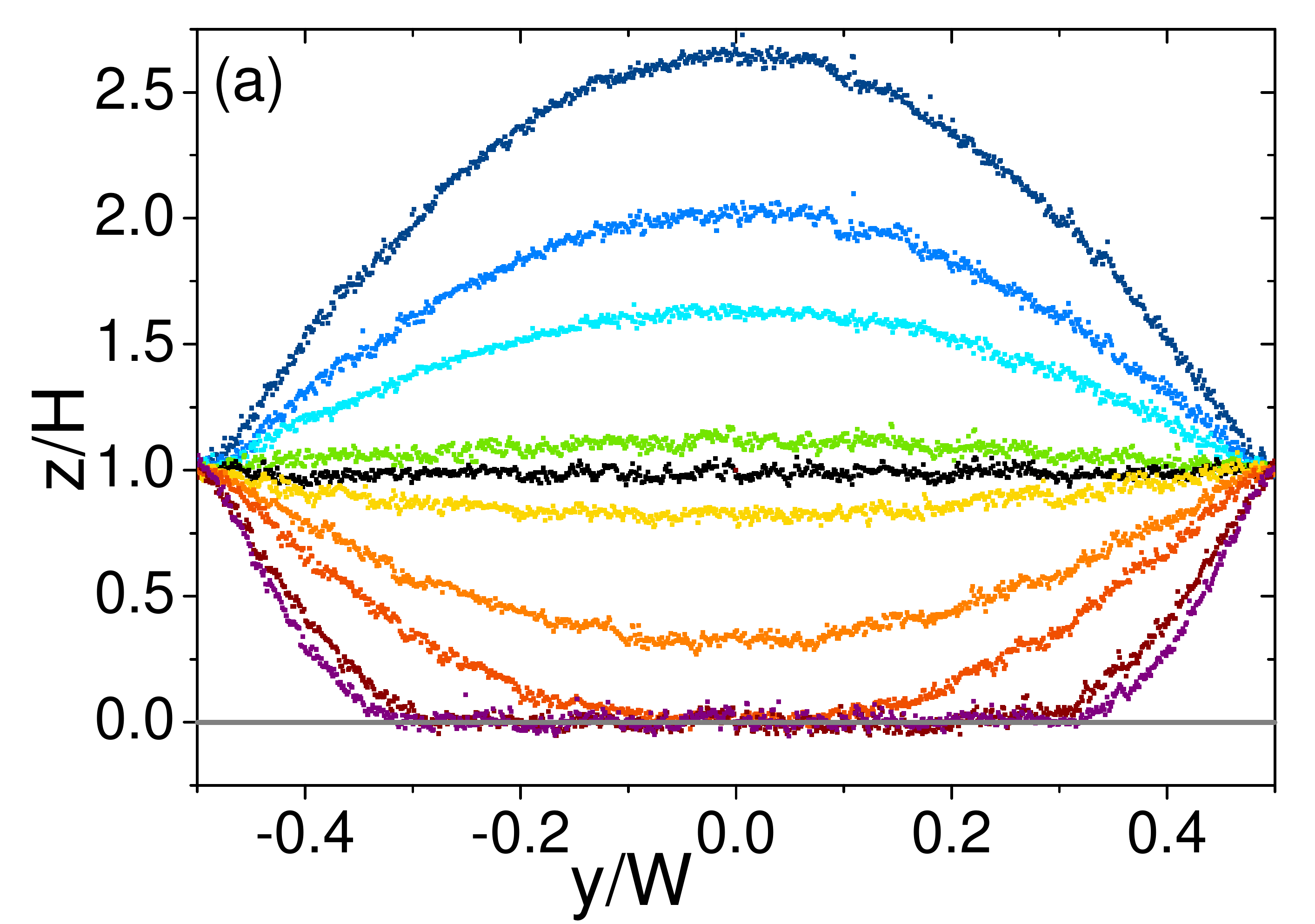}
\includegraphics[height=4.75cm]{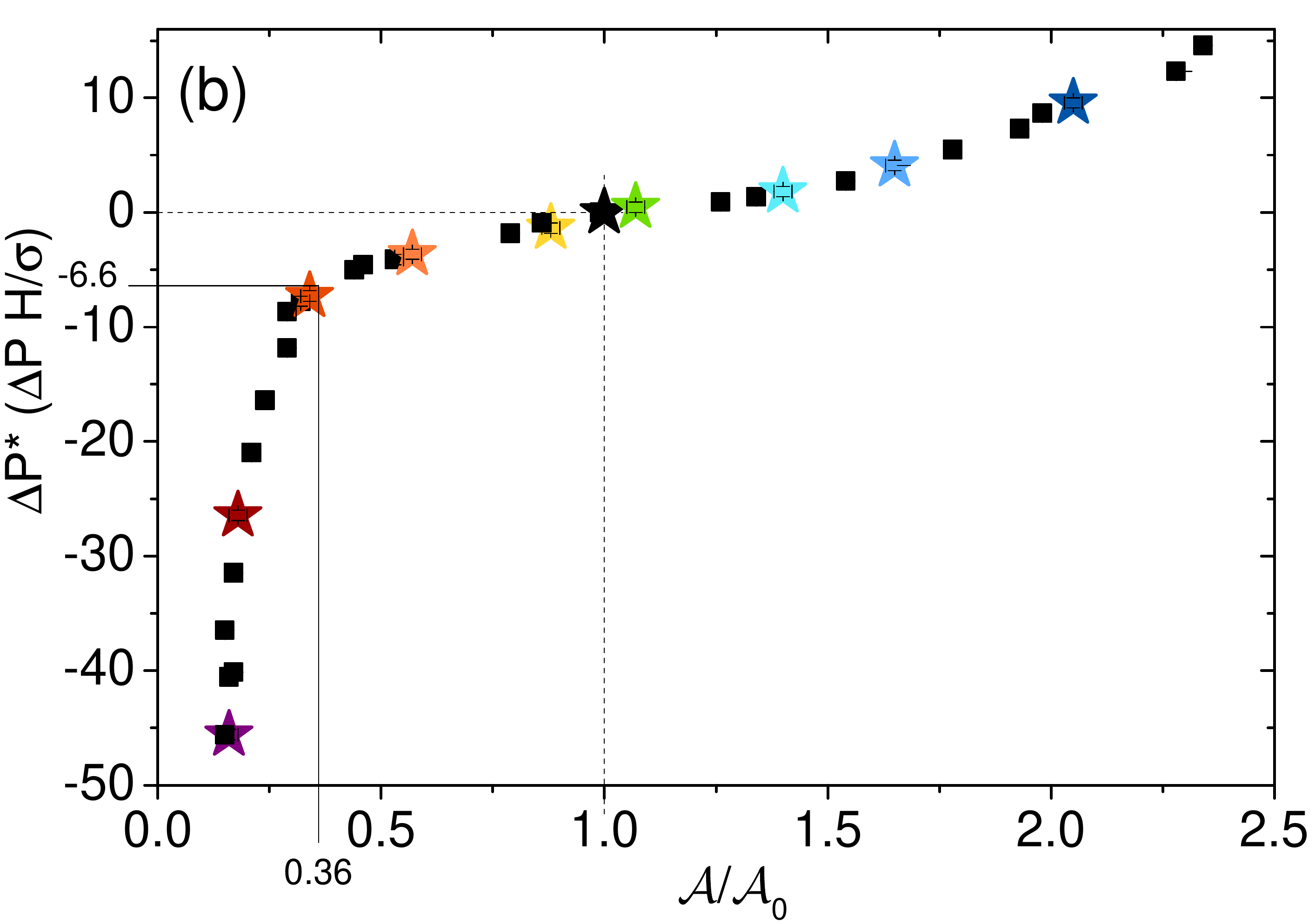}
\includegraphics[height=4.75cm]{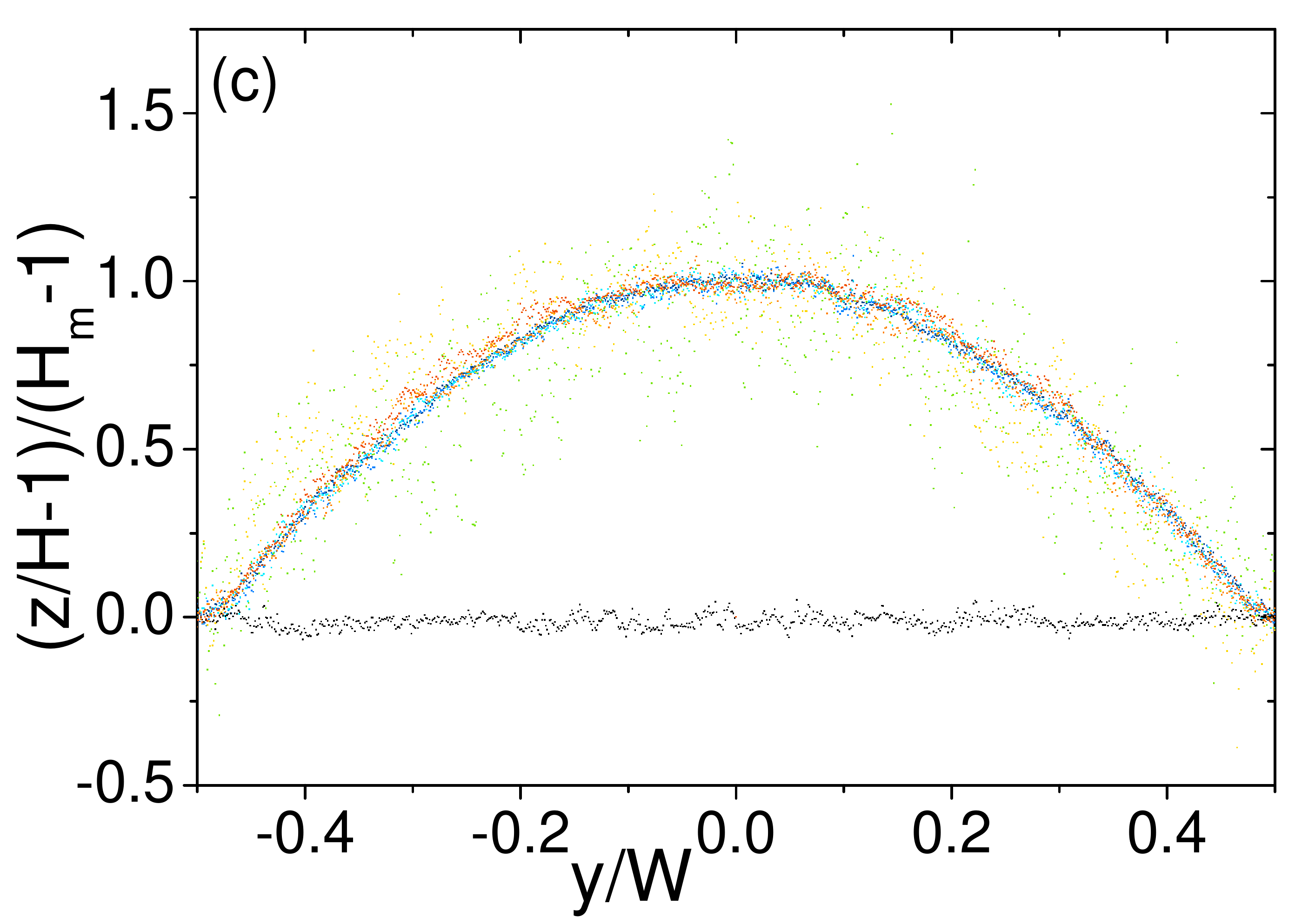}
\includegraphics[height=4.75cm]{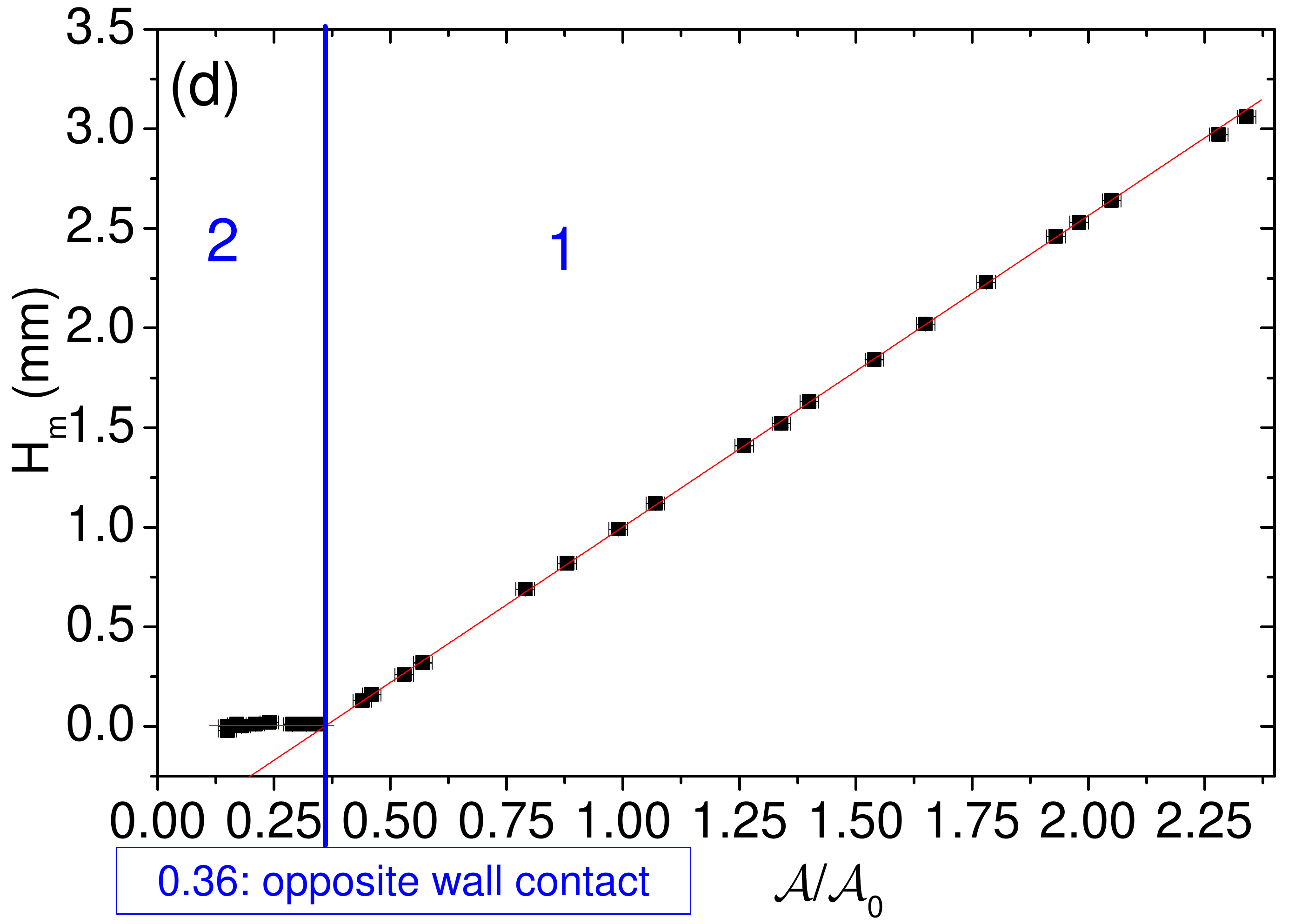}
\caption{(\textit{a}) Series of membrane profiles for (from top to bottom) decreasing values of the transmural pressure $\Delta P$. The profiles have been normalized by the channel width in the $y$ direction and undeformed depth in the $z$ direction. The horizontal grey line shows the position of the channel bottom. (\textit{b}) Dimensionless transmural pressure $\Delta P^*$ across the membrane as a function of the level of collapse. Dotted lines indicate the undeformed configuration of the channel ($\Delta P^*=0$), while the thin lines indicate the point of opposite wall contact. The pentagrams give the ($\Delta P^*,\mathcal{A}/\mathcal{A}_0$) value of the profiles presented in panel (\textit{a}) (same colours are used). (\textit{c}) Same profiles as above for $\mathcal{A}/\mathcal{A}_0>0.36$, rescaled by their maximum deflection. Note that the reference profile (neutral point) has been shifted to $z=0$. The rescaling also amplifies the larger relative scatter for small deflections. (\textit{d}) Height $H_m$ in the middle of the channel cross-section ($y=0$) as a function of the level of collapse. The intersection of the two regimes accurately defines the point of opposite wall contact.\label{fig:tube-law} }
\end{figure}
\end{center}

The shape of the channel static cross-section for various levels of uniform loading is shown in figure~\ref{fig:tube-law} (\textit{a}). The channel is inflated under positive transmural pressure ($P_{in}>P_{out}$), and collapsed under negative transmural pressure. We quantify the level of collapse (or inflation) of the channel by its dimensionless cross-sectional area $\mathcal{A}/\mathcal{A}_0$, where $\mathcal{A}_0=WH$ is the cross-sectional area at zero transmural pressure. $\mathcal{A}/\mathcal{A}_0=1$ is the neutral point of the system, increasing values of $\mathcal{A}/\mathcal{A}_0>1$ correspond to increasing inflation of the membrane and decreasing $\mathcal{A}/\mathcal{A}_0<1$ measures increasing collapse of the channel. The values of $\mathcal{A}/\mathcal{A}_0$ were measured by fitting a polynomial function (tenth order) to the membrane profiles and integrating it over the width of the channel. The high degree of the polynomial was chosen to ensure that even the most collapsed profiles could be fitted. The experimental resolution on all values of $\mathcal{A}/\mathcal{A}_0$ is 0.02.\\

The transmural pressure $\Delta P$ (scaled by the capillary pressure $\sigma /H$ over the undeformed channel depth) is shown as a function of the level of collapse in figure~\ref{fig:tube-law} (\textit{b}). It is qualitatively similar to the tube law relating the transmural pressure to the cross-sectional area of an elastic tube (as measured for instance by~\citet{juel2007reopening}), with a quasi-linear evolution around the neutral point, a very steep reduction in pressure once the membrane has come close to the bottom of the channel and a convex growth for positive transmural pressure. The behaviour of the elasto-rigid channel significantly differs from the tube law under inflation only: once a tube is inflated, $\Delta P$ grows very rapidly because any further increase in the tube cross-sectional area requires large stretching of the tube wall, whereas comparatively lesser stretching of the flat membrane is necessary to increase the channel cross-sectional area, resulting in a lesser gradient in $\Delta P$. As in the experiments performed with elastic tubes, no dewetting of the oil occurs either on the Perspex or the latex, so that even in the most collapsed channel configurations a thin layer of oil is left between the channel bottom and the membrane. For simplicity, we call \textit{point of first opposite wall contact} (POWC) the critical level of collapse $(\mathcal{A}/\mathcal{A}_0)_{\mathrm{c}}$ for which the thickness of that oil layer becomes smaller than our experimental resolution. We then estimate the layer to be less than 50 $\mu$m thick. We refer to levels of collapse such that $1>\mathcal{A}/\mathcal{A}_0>(\mathcal{A}/\mathcal{A}_0)_{\mathrm{c}}$ as low levels of collapse, and to those for which $\mathcal{A}/\mathcal{A}_0<(\mathcal{A}/\mathcal{A}_0)_{\mathrm{c}}$ as high levels of collapse.  \\

Despite a qualitatively similar constitutive behaviour, the evolution of the cross-sectional profiles as the transmural pressure decreases is considerably simplified in the elasto-rigid channel compared to a tube. For low levels of collapse, all cross-sectional shapes of the elasto-rigid channel are similar, as demonstrated by the collapse of the data onto a single curve after rescaling the profiles by their maximum deflection from the neutral line (see figure~\ref{fig:tube-law} (\textit{c})). Those membrane profiles can be approximated by a family of quadratic functions, with a single scaling parameter dependent on $\mathcal{A}/\mathcal{A}_0$ (see appendix~\ref{sec:appendixA} for plots of the profiles and their fits). For high levels of collapse, the width of the centrally collapsed region of near opposite wall contact increases, approximately linearly, with the level of collapse. This simple behaviour contrasts with the series of shapes adopted by an elastic tube as the level of collapse increases: in the elasto-rigid channel, the measurement of a single parameter such as the level of collapse $\mathcal{A}/\mathcal{A}_0$ is  enough to fully characterise the evolution of the cross-sectional profile among a family of self-similar shapes for low levels of collapse. \\
The self-similarity of the channel cross-sectional profiles for low levels of collapse also allows us to define very precisely the POWC by plotting the membrane height in the middle of the cross-section $H_m$ as a function of $\mathcal{A}/ \mathcal{A}_0$ (see figure~\ref{fig:tube-law} (\textit{d})). Two regimes can be identified: for low collapse, $H_m$ scales linearly with $\mathcal{A}/ \mathcal{A}_0$ and the membrane profiles can be rescaled as described above (regime (1) in figure~\ref{fig:tube-law} (\textit{d})); for high collapse, the membrane has come close to the bottom of the channel and $H_m=0$ (regime (2)). The intersection of the two regimes gives an accurate measurement of the POWC for our elasto-rigid channel, which occurs for $\mathcal{A}/ \mathcal{A}_0=0.36 \pm 0.02$. \\

\subsection{Relevant dimensionless parameters}
\label{sec:param}
The high level of control on the cross-sectional shape of our elasto-rigid channel and the simplicity of the profiles adopted by the membrane respectively for low or high levels of collapse allow us to perform an extensive study of the influence of the level of initial collapse, specified by $\mathcal{A}_i/ \mathcal{A}_0$ where $\mathcal{A}_i$ is the cross-sectional area of the channel prior to air injection, on the propagation of the air finger. In terms of fluid parameters, we focus on the dependence of the finger propagation mode on the capillary number $\mathrm{Ca}=\mu U/\sigma$ (with $U$ the finger speed, $\mu$ the dynamic viscosity of the fluid and $\sigma$ the surface tension of the fluid), which is the essential control parameter in viscous fingering problems. The ratio of gravitational forces to surface tension forces, quantified by the Bond number $Bo=\frac{\rho g H^2}{4\sigma}$ is $Bo=0.11$. In a rigid channel, buoyancy does not qualitatively affect the finger shape for $Bo<1$~\citep{jensen1987effect}; we thus expect gravity to be negligible in our experiments. The Reynolds number $Re=\frac{\rho U H}{\mu}$ is of order 1 for the highest finger speed we considered. \\

\section{Results}
\subsection{Effect of the level of initial collapse}
\label{sec:collapse}
We investigated the effect of the static cross-section of the channel on the air finger propagation mode at a fixed value of the capillary number, $\mathrm{Ca}=0.47\pm 0.02$. In a rigid-walled Hele-Shaw channel, this value of the capillary number is high enough for the finger shape to be close to the half-width finger observed in the limit of negligible surface tension by~\citet{saffman1958penetration}. We will see in this section that much wider fingers can be observed in a compliant channel at such high values of $\mathrm{Ca}$. \\

\subsubsection{Low levels of initial collapse}
\label{sec:lowColl}
\begin{figure}
\begin{center}
\includegraphics[height=1.4cm]{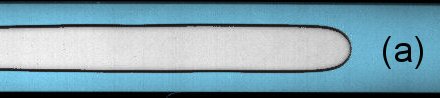}
\includegraphics[height=1.4cm]{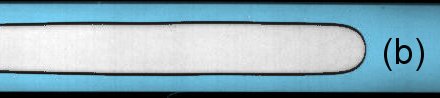}\\
\includegraphics[height=1.4cm]{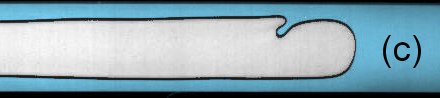}
\includegraphics[height=1.4cm]{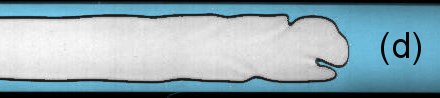}\\
\includegraphics[height=1.4cm]{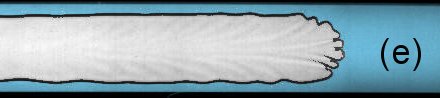}
\includegraphics[height=1.4cm]{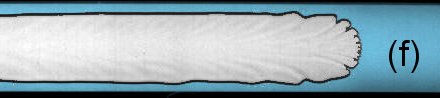}
\caption{Shape of the finger for, from (\textit{a}) to (\textit{f}), increasingly collapsed initial conditions. The corresponding values of $\mathcal{A}_i/ \mathcal{A}_0$ are: (\textit{a}) 1.01, (\textit{b}) 0.87, (\textit{c}) 0.70, (\textit{d}) 0.54, (\textit{e}) 0.43 and (\textit{f}) 0.37. \label{fig:finger_low_collapse}}
\end{center}
\end{figure}

The shapes of the fingers obtained at $\mathrm{Ca}=0.47$ for increasing levels of low initial collapse, from the undeformed channel to the vicinity of the POWC, are shown in figure~\ref{fig:finger_low_collapse}. For initially flat or only very slightly collapsed membrane shapes, the finger profile qualitatively resembles the classical Saffman-Taylor finger observed when injecting air into an oil-filled fully rigid Hele-Shaw channel (see figures~\ref{fig:finger_low_collapse} (\textit{a},\textit{b})). In a rigid-walled channel, the finger profile is set by the balance of viscous and surface tension forces through the Young-Laplace equation, which relates the curvature of the interface to the pressure jump across it. The curvature is maximum at the finger tip where the pressure difference is largest between the gas and the flowing liquid and decreases to zero far behind the tip where the fluid is at rest. The resulting profile sets the finger width and is qualitatively similar to what we observe in the elasto-rigid channel if the membrane is initially flat. As the level of collapse increases, the finger tip becomes unstable (figure~\ref{fig:finger_low_collapse} (\textit{c}, \textit{d})) and develops a secondary fingering instability (figure~\ref{fig:finger_low_collapse} (\textit{e}, \textit{f})). This small scale pattern only grows to a finite amplitude and travels with the main finger tip, in a similar manner to the fingering instability observed in an elasto-rigid radial Hele-Shaw cell~\citep{pihler2012suppression}. In rigid-walled Hele-Shaw channels, the classical Saffman-Taylor finger profile is linearly stable~\citep{bensimon1986stability} but can be destabilised by the introduction of local perturbations which modify its curvature, resulting in the growth of an additional fingering pattern~\citep{couder1986narrow}. A similar coupling between the profile of the main finger and the small scale fingering pattern at the finger tip may occur in our elasto-rigid channel. Here we focus solely on the propagation of the main finger, and refer the reader to~\citet{ducloue2016fingering} for a detailed study of the secondary fingering instability in a compliant channel.\\

\begin{figure}
\begin{center}
\includegraphics[height=5cm]{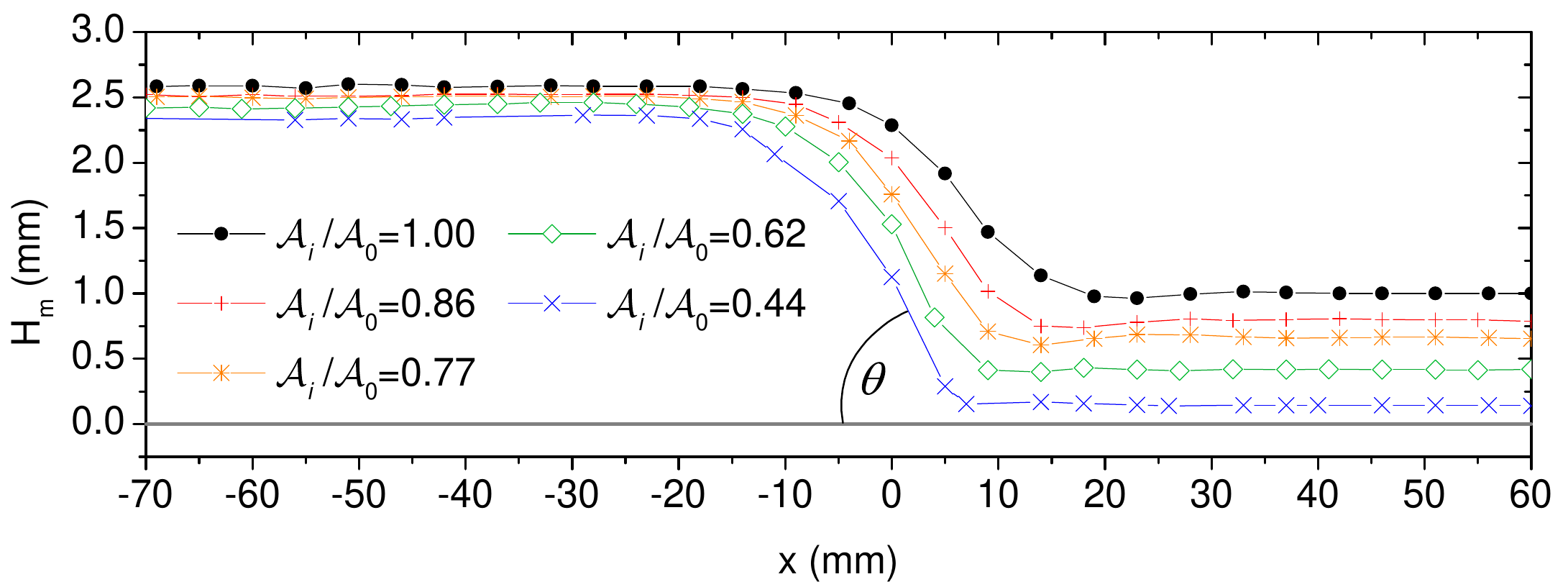}
\end{center}
\caption{Axial profile of the membrane (height $H_m$ in the middle of the cross-section versus position in the direction of propagation) during reopening by the air finger for various levels of initial collapse (corresponding values of $\mathcal{A}_i/ \mathcal{A}_0$ are given in the legend). The finger tip identified on the images is sitting at $x=0$. The $x$-position has been reconstructed from the time series of images acquired at a fixed position by using the constant finger speed. The reopening angle $\theta$ is defined by $\tan \theta=\frac{\mathrm{d}H_m}{\mathrm{d}x}$, and hence $\theta$ is negative because the channel walls are converging. \label{fig:axial_profile_low_collapse}}
\end{figure}

\paragraph{\textbf{Case of $\mathbf{\mathcal{A}_i}/\mathbf{\mathcal{A}_0=1}$}:}
\label{sec:undeformed}
The similarity of the finger profile to the classical Saffman-Taylor finger may be surprising considering that the elasto-rigid channel inflates as the finger propagates: the membrane is lifted by more than the channel's initial depth far behind the finger tip, as shown in figure~\ref{fig:axial_profile_low_collapse} by the height profile $H_m(x)$ half-way across the channel width. Most of this lifting however is done by a liquid wedge propagating ahead of the air interface, so that the channel is already almost fully inflated at the finger tip. Therefore the finger tip experiences an effective rigid channel which local aspect ratio sets the curvature of the interface. A similar regime was observed in two- and three-dimensional models of the reopening of elastic tubes: below a critical capillary number dependent on the elastic properties of the tube (2D models) or the level of initial collapse (3D model), the tube inflated ahead of the air finger, and the reopening finger could be modelled as propagating in an equivalent rigid inflated geometry, with very good agreement with computational results~\citep{gaver1996steady, jensen2002steady, hazel2003three}. The distance between the region where the membrane deforms in response to the pressure in the liquid wedge and the finger tip behind it uncouples the finger curvature from the axial changes in the channel cross-section, explaining why the propagating finger resembles a Saffman-Taylor finger. \\

\begin{figure}
\begin{center}
\includegraphics[height=5.0cm]{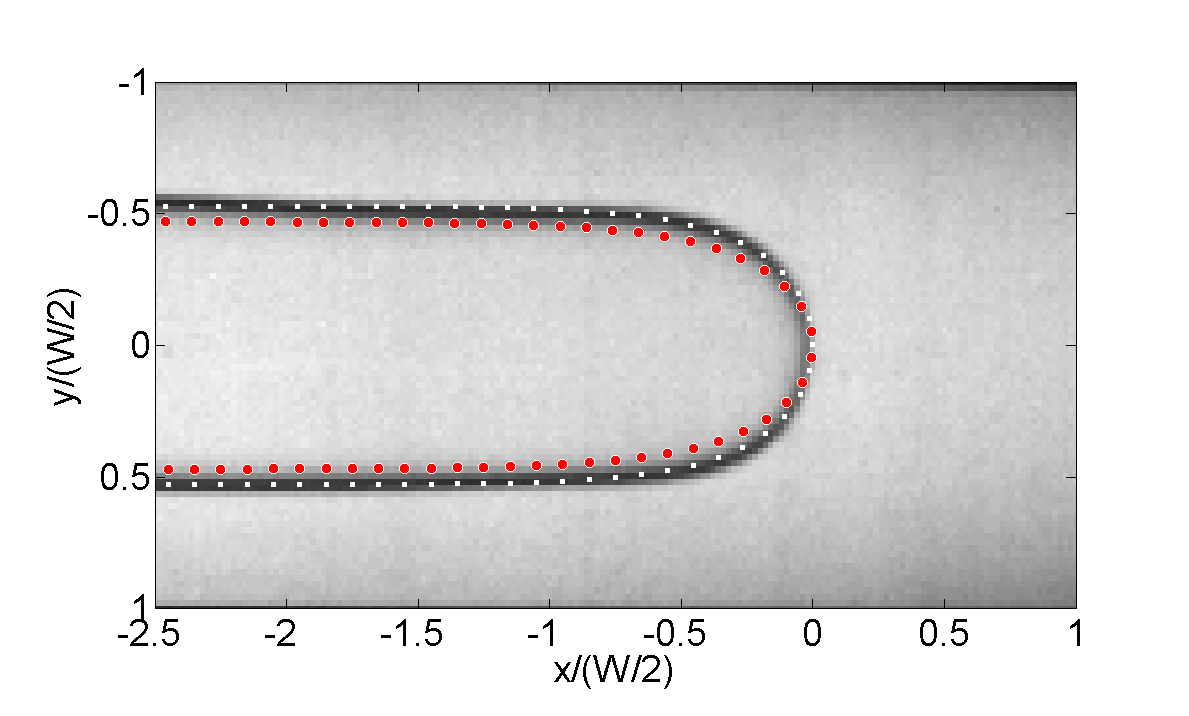}
\end{center}
\caption{Enlargement of the experimentally observed finger tip for an initially undeformed membrane ($\mathcal{A}_i/ \mathcal{A}_0$=1.01). The red dots show the profile computed by~\citet{saffman1958penetration} for $\lambda=0.47$, which would be the expected width of a finger propagating at the same modified capillary number in a fully rigid channel. The white solid squares show the same profile computed for $\lambda=0.53$, which takes into account the effective aspect ratio of the inflated channel at the finger tip.\label{fig:comp-ST}}
\end{figure}

An enlargement of the finger tip is presented in figure~\ref{fig:comp-ST}. The width of the finger relative to the channel width, $\lambda$, is measured to be $0.51\pm0.01$. A quantitative comparison with the experimental finger width obtained in a fully rigid Hele-Shaw channel can be made using the data of~\citet{tabeling1987experimental}, who measured $\lambda$ as a function of a modified capillary number $1/B=12\alpha^2\mathrm{Ca}$, where $\alpha$ is the aspect ratio of the channel. Using $\alpha=W/H=30$ for our undeformed channel leads to $1/B=5100$ for the profile presented in figure~\ref{fig:comp-ST}. At this value, the finger propagating in the rigid channel has a relative width $\lambda=0.47\pm0.01$ as measured by~\citet{tabeling1987experimental}, which is 8\% narrower than observed in the elasto-rigid channel. The profile computed by~\citet{saffman1958penetration} for the finger tip is in excellent agreement with the experimental finger shapes of ~\citet{tabeling1987experimental} close to the half-width limit. This profile is plotted for $\lambda=0.47$ (red dots) in figure~\ref{fig:comp-ST}. The experimental finger shape has a lower tip curvature than the computed profile: this is because of the inflation of the membrane ahead of the finger in the experiment. Using the data from the reopening profile in figure~\ref{fig:axial_profile_low_collapse}, we can calculate that the finger tip propagates in a channel of effective aspect ratio 13, corresponding to a modified capillary number $1/B \sim 1000$. In a rigid channel, the finger width measured for this lower value of the modified capillary number is $\lambda=0.53\pm0.02$~\citep{tabeling1987experimental}, which is a better estimate for our experimental data, as shown by the white squares in figure~\ref{fig:comp-ST}. \\

\begin{figure}
\begin{center}
\includegraphics[height=5cm]{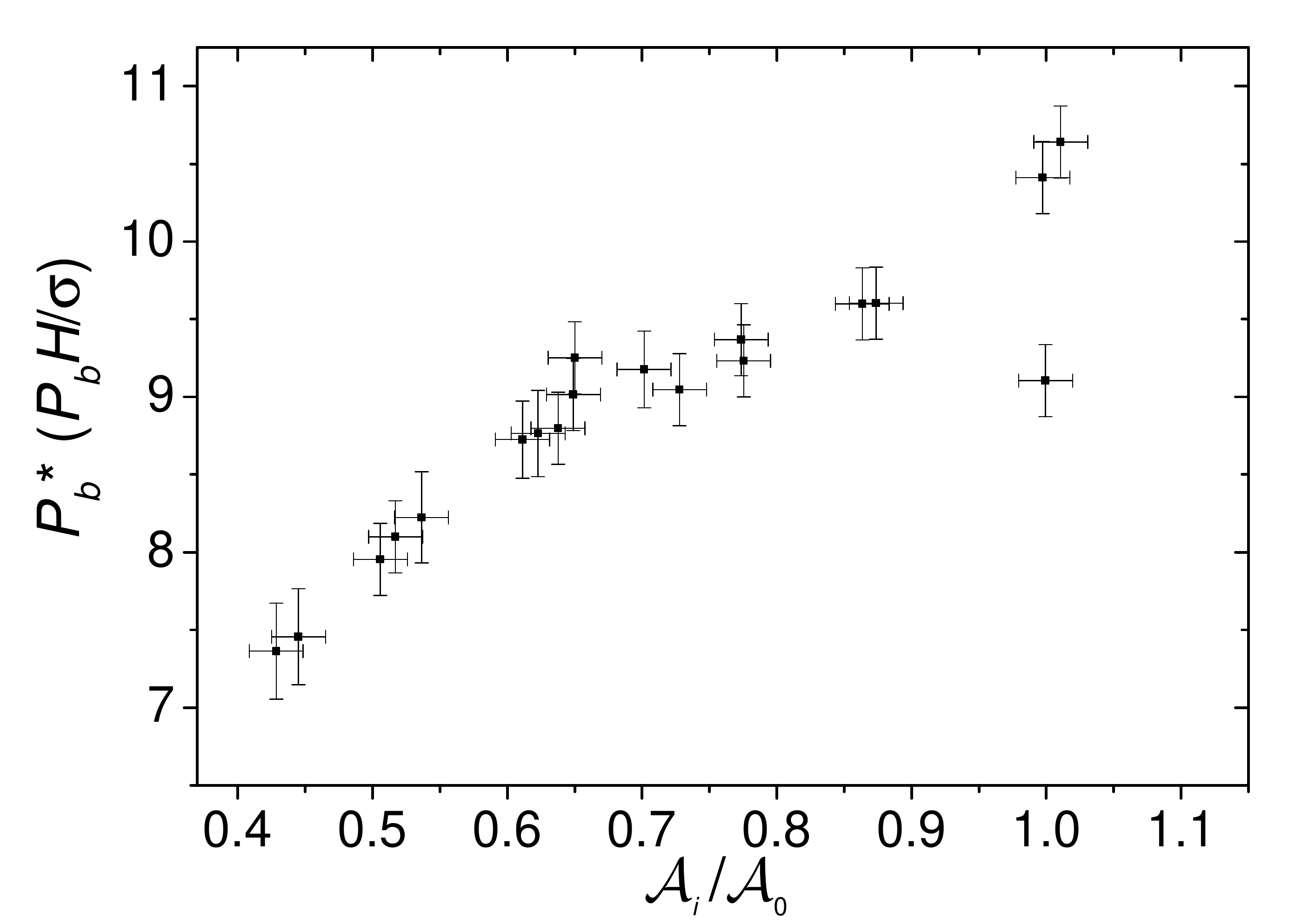}
\caption{Dimensionless air finger pressure, as a function of the level of initial collapse. \label{fig:pressure_low_collapse}}
\end{center}
\end{figure}

\paragraph{\textbf{Case of $\mathbf{\mathcal{A}_i}/\mathbf{\mathcal{A}_0<1}$}:}
As the channel is gradually collapsed, the pressure required to drive the finger at the prescribed capillary number $\mathrm{Ca}=0.47$ decreases, as can be seen in figure~\ref{fig:pressure_low_collapse}: it is easier to reopen the channel from a more collapsed initial configuration. This behaviour arises because of the nature of the flow in the compliant channel: the fluid layer far ahead of and far behind the propagating air finger is at rest and gets redistributed at the finger tip. The higher the level of initial collapse, the lesser the volume of fluid which gets redistributed, which minimises the amount of work required to propagate the finger as observed by~\citet{hazel2003three} in their numerical simulations of the reopening of elastic tubes for low levels of initial collapse. \\
Consistently with the lower pressure in the reopening finger, the final reopening height of the channel decreases for higher levels of initial collapse, as can be seen in figure~\ref{fig:axial_profile_low_collapse}. It can be noticed that the decrease in reopening height for increasing levels of initial collapse is smaller than the change in initial height. This is due to the non-linearity of the channel constitutive behaviour (see figure~\ref{fig:tube-law} (\textit{b})) in which the membrane gets stiffer for large inflated states than for moderately collapsed (or inflated) configurations. \\

\begin{figure}
\begin{center}
\includegraphics[height=5.5cm]{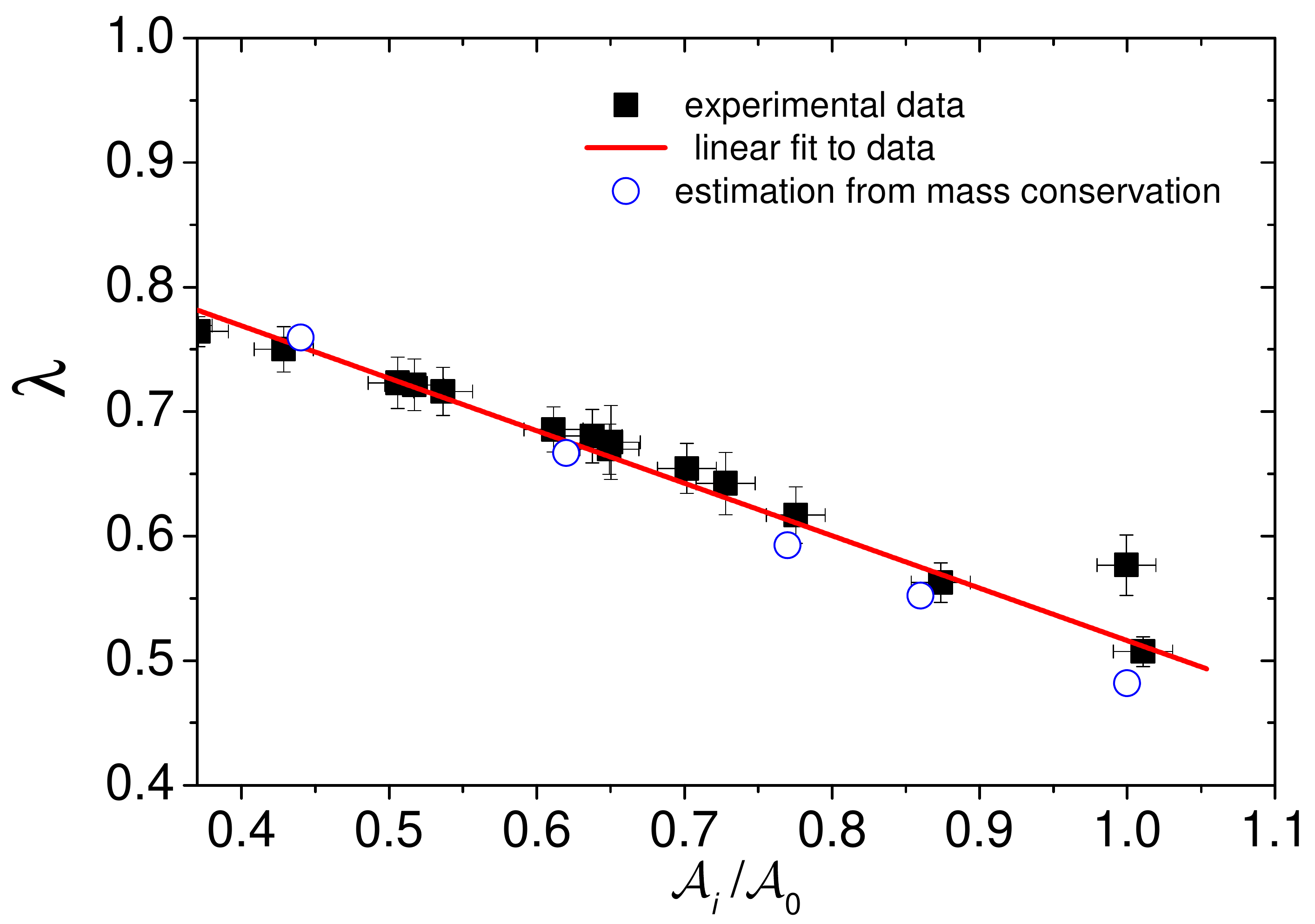}
\caption{Relative width of the finger $\lambda$ as a function of the initial level of collapse. The red line is a linear least-squares fit to the experimental data. Blue empty circles are estimates of $\lambda$ from a simplified mass conservation argument (equation~\ref{eq:mass}).\label{fig:lambda_low_collapse}}
\end{center}
\end{figure}

As the level of collapse increases, the finger shape deviates further from the finger width expected in rigid channels: the thinner liquid layer and the consequent larger aspect ratio at the finger tip in more collapsed geometries would promote narrower fingers in fully rigid channels. On the contrary, the relative width of the finger $\lambda$ increases linearly in our elasto-rigid channel, up to $\lambda=0.75$ close to the POWC, as quantified in figure~\ref{fig:lambda_low_collapse}. However, a single centred symmetric finger is still observed, even though the channel initial cross-section becomes increasingly non-uniform (the depth in the middle of the cross-section is only 15\% of the undeformed depth for $\mathcal{A}_i/ \mathcal{A}_0=0.44$). In a rigid channel, \citet{franco2016sensitivity} have shown that centred, step-like occlusions of width $W/4$ lead to asymmetric fingers for occlusions as low as a few percent of the channel height at the modified capillary number we are considering. The moderately collapsed membrane takes a smoother shape than a step-like occlusion, but we expect that minimisation of viscous resistance would promote off-centred fingers in a rigid channel of the same cross-sectional shape. In a compliant vessel, because of the reopening of the membrane to a uniform or inflated cross-section over a short distance ahead of the finger tip, only a small volume of liquid, located in the wedge at the finger tip, flows in a centrally constricted channel. This lessens the impact of the greater viscous resistance in the middle of the compliant channel compared to a rigid channel with the same cross-section. However, minimisation of viscous resistance still promoted asymmetric fingers for the reopening of moderately collapsed elastic tubes at high capillary numbers~\citep{heap2009bubble}. The existence of those fingers is dependent on the shape taken by the tube cross-section as it reopens, which is set by the tube law. The constitutive behaviour of our elasto-rigid channel imposes that it reopens to an inflated state ahead of the finger tip for all levels of low initial collapse at $\mathrm{Ca}=0.47$ (see figure~\ref{fig:axial_profile_low_collapse}). Our channel cannot therefore sustain steadily propagating asymmetric fingers, and we observe a single propagation mode for low levels of initial collapse at fixed $\mathrm{Ca}=0.47$. This symmetric round-tipped finger is analogous to the symmetric peeling air finger described in the reopening of an elastic tube~\citep{juel2007reopening}.\\

The increase in the finger width with the level of initial collapse can be predicted by a simple mass conservation argument: $\lambda$ can be estimated by considering that the area of the initial cross-section $\mathcal{A}_i$ far ahead of the air finger should be equal to the wetted area of the final cross-section $\mathcal{A}_f$ far behind the finger tip. Assuming that the volume of the films left on the top and bottom walls is negligible compared to the amount of liquid at the sides of the finger and that films of uniform thickness and average height $\mathcal{A}_f/W$ separate the finger from the side walls, the width of the finger then satisfies 
\begin{equation}
\lambda=1-\mathcal{A}_i/\mathcal{A}_f
\label{eq:mass}
\end{equation}
This result is plotted with empty circles in figure~\ref{fig:lambda_low_collapse} and is in good agreement with the experimental data, with the correct linear trend and a slight deviation at lower levels of collapse. The deviation is likely to be due to the thin films left on the membrane and the channel bottom after the propagation of the finger, which we have neglected in this estimation but get thicker for larger channel initial depth ahead of the finger (as we have experimentally observed but not quantified, and has been calculated by~\citet{park1984two} for a Saffman-Tayor finger propagating in a rigid-walled Hele-Shaw channel). \\

\begin{figure}
\begin{center}
\includegraphics[height=5.5cm]{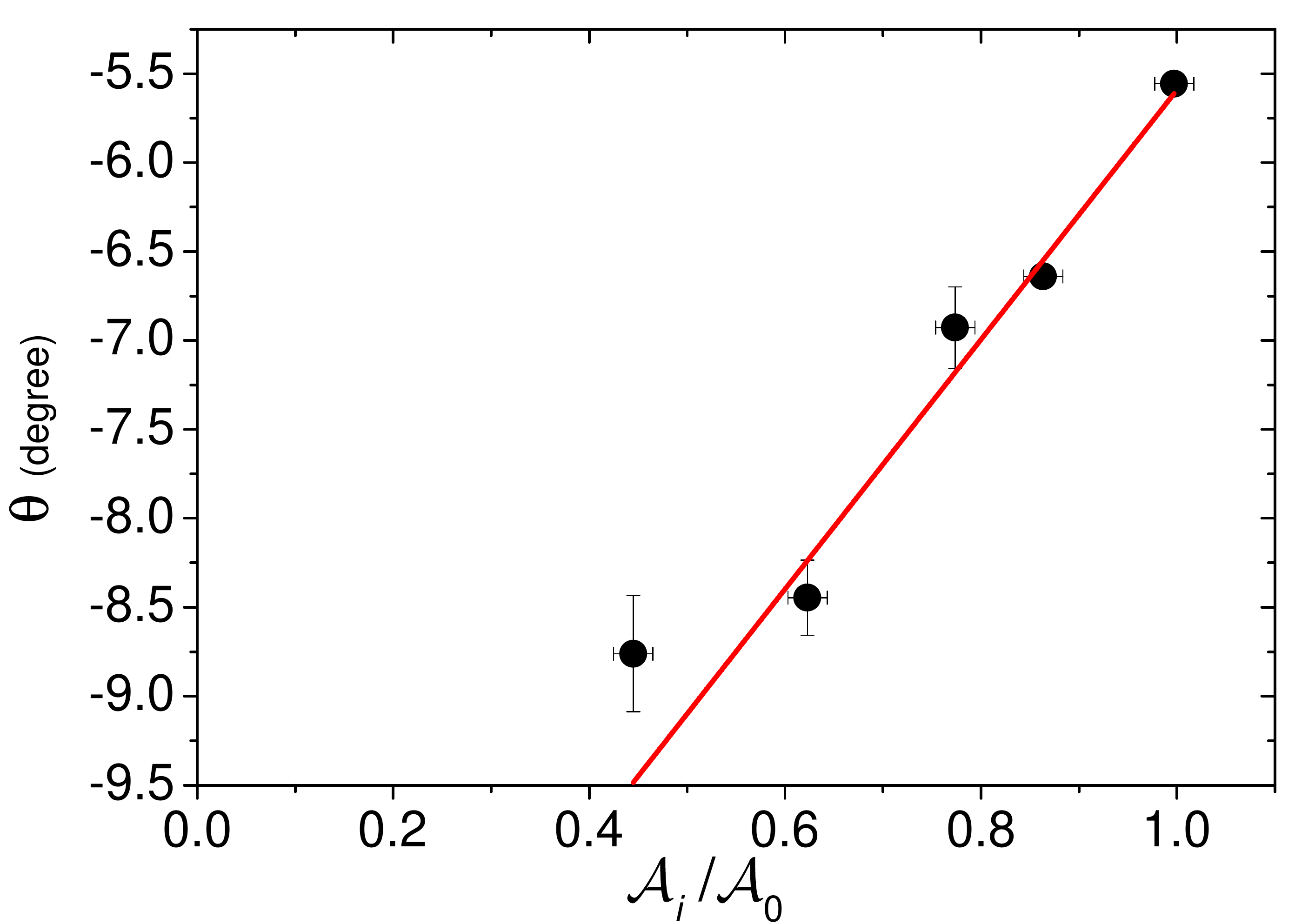}
\caption{Angle $\theta$ of the taper at the reopening front, measured from the profiles given in figure~\ref{fig:axial_profile_low_collapse}, as a function of the level of initial collapse. The red line is a linear least-squares fit to the experimental data. \label{fig:alpha_low_collapse}}
\end{center}
\end{figure}

The membrane reopens with a steeper angle as the level of collapse increases (see figure~\ref{fig:axial_profile_low_collapse}). 
The combination of a greater reopening angle and smaller volume of fluid to redistribute at the finger tip imposes that the length of the liquid wedge at the tip is reduced as the level of initial collapse increases, as can indeed be seen from the reopening profiles in figure~\ref{fig:axial_profile_low_collapse}. As a consequence, the finger tip approaches the region of imposed cross-section $\mathcal{A}_i$ and propagates in a strongly convergent geometry. The angle $\theta$ of the locally tapered channel experienced by the tip, measured by computing the derivative of the reopening profiles presented in figure~\ref{fig:axial_profile_low_collapse}, is plotted as a function of the level of initial collapse in figure~\ref{fig:alpha_low_collapse}. This angle is negative, reflecting the fact that the finger propagates in a locally convergent cell, and its absolute value increases approximately linearly with the level of initial collapse.\\
In a fully rigid rectangular channel, such a convergent taper was found to cause the propagation of fingers of higher tip curvature compared to uniform cells~\citep{zhao1992perturbing}. However, the authors pointed out that finger narrowing in their rigid-walled channel was due to increased fluid velocity in the thinning gap over the finger tip length. This increased velocity is imposed by mass conservation in the rigid cell because the fluid is flowing in the whole channel ahead of the finger. By contrast, in the compliant vessel, the finger propagates steadily in a convergent geometry which is stationary in the frame of reference of the finger tip. \\
The convergent taper promotes greater restoring capillary forces and weaker destabilising viscous forces as shown in the linear stability analysis of a flat interface propagating in a rigid tapered Hele-Shaw channel~\citep{al2013controlling}. For the fully developed finger, the convergence of the cell at the finger tip will create higher resistance to propagation in the direction of the taper, that is, in the direction of the channel axis. Cross-sectional variations in the resistance to finger propagation have been modelled in rigid rectangular channels by introducing anisotropic surface tension~\citep{amar1993viscous,combescot1994saffman}. In this framework, the higher viscous resistance to propagation along the centreline of the tapered channel is equivalent to a higher surface tension in this direction. Calculations of the finger width in the limit of low surface tension and for small anisotropy show that such adverse anisotropic surface tension is expected to select a wider finger, of smaller tip curvature, compared to the case of isotropic surface tension. The reopening angle of the compliant channel may set the curvature at the finger tip via a similar mechanism. \\

\subsubsection{High levels of initial collapse}
\label{sec:highColl}
The constitutive behaviour of our elasto-rigid channel resulted in the observation of a single propagation mode for low levels of initial collapse, a round-tipped symmetric finger. We now turn to high levels of collapse, for which a variety of propagation modes were observed during the reopening of an elastic tube, including the peculiar pointed finger for very strong initial collapse. \\

\begin{figure}
\begin{center}
\includegraphics[height=1.4cm]{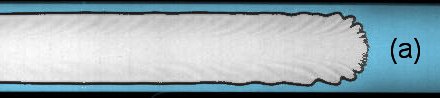}
\includegraphics[height=1.4cm]{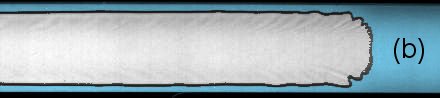}\\
\includegraphics[height=1.4cm]{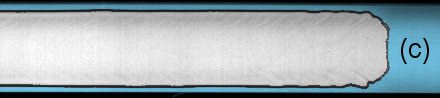}
\includegraphics[height=1.4cm]{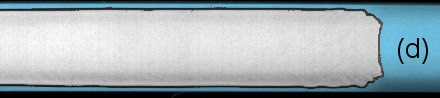}
\caption{Shape of the finger for strongly collapsed initial conditions. The corresponding values of $\mathcal{A}_i/\mathcal{A}_0$ are (\textit{a}) 0.31, (\textit{b}) 0.25, (\textit{c}) 0.21 and (\textit{d}) 0.16. \label{fig:finger_high_collapse}}
\end{center}
\end{figure}

\begin{figure}
\begin{center}
\includegraphics[height=5cm]{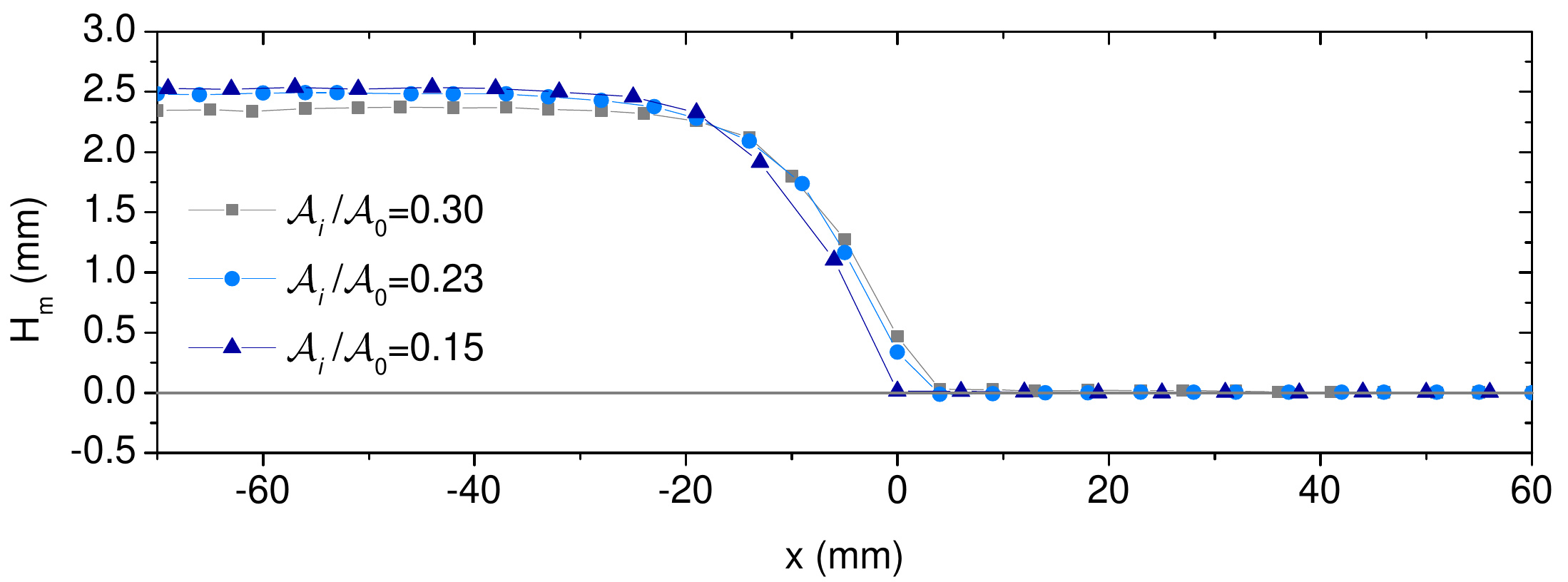}
\caption{Axial reopening profiles in the middle of the cross-section (height $H_m$ versus position in the direction of propagation) for various levels of high initial collapse (corresponding values of $\mathcal{A}_i/ \mathcal{A}_0$ are given in the legend). \label{fig:profiles_high_collapse}}
\end{center}
\end{figure}

For high initial collapse, the shape of the channel cross-section qualitatively changes: only a very thin layer of liquid is left between the membrane and the channel bottom in the centre of the cross-section, and the width of this highly collapsed region increases with the level of collapse. The finger profiles obtained at $\mathrm{Ca}=0.47\pm0.02$ for increasing degrees of high initial collapse are illustrated in figure~\ref{fig:finger_high_collapse}.
A single centred symmetric finger is still observed as the width of the collapsed region increases but the tip of that finger undergoes a qualitative evolution: it flattens in the centre of the channel and right behind this flat front the finger widens over a short distance to the constant finger width expected far behind the tip. For very high levels of initial collapse (as in figure~\ref{fig:finger_high_collapse} (\textit{c}), (\textit{d})), the front of the finger is not perfectly flat but slightly bent inwards in its centre part, which we attribute to small differences in the thickness of the liquid layer in the collapsed region (these are below our experimental resolution but suggested by smoothing of the experimental cross-sectional profiles). The flat tip of the finger infiltrates the thin liquid layer in the middle of the channel, as confirmed by the reopening profiles in figure~\ref{fig:profiles_high_collapse}: the liquid wedge at the finger tip becomes very small due to the minimal amount of liquid redistributed, and the finger tip reaches into the thin layer of oil. This finger exhibits the same counter-intuitive behaviour as the pointed finger by propagating in the region of the channel where viscous resistance is maximum. We shall see later in this section that the flat-tipped finger is the same propagation mode as the pointed finger observed in strongly collapsed tubes. It also exhibits a discontinuous transition towards another propagation mode with decreasing $\mathrm{Ca}$, which is reported in section~\ref{sec:Ca}. \\

\begin{figure}
\begin{center}
\includegraphics[height=5.5cm]{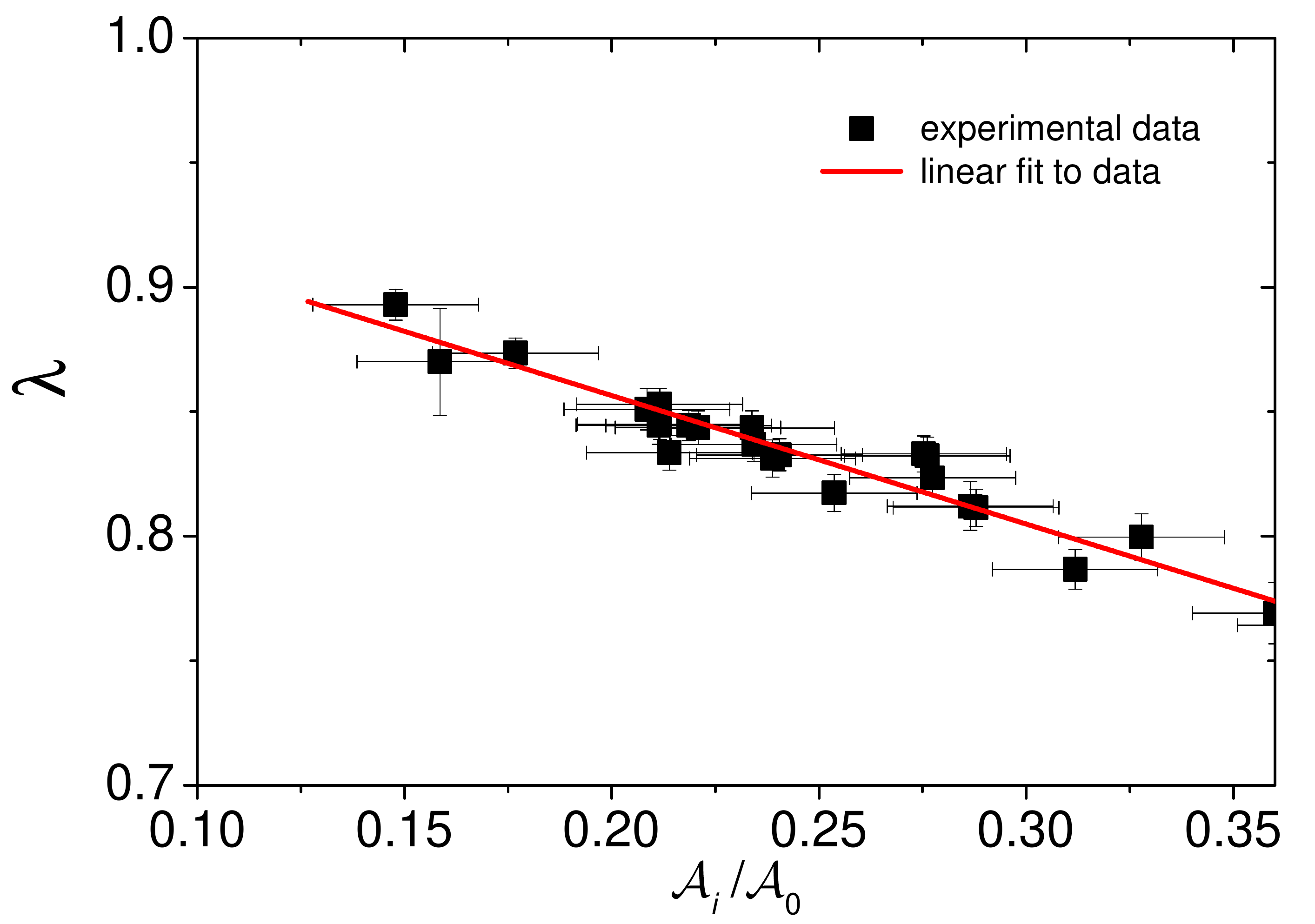}
\caption{Relative width of the finger $\lambda$ as a function of the level of initial collapse $\mathcal{A}_i/\mathcal{A}_0$ for strong initial collapse. The red line is a linear least-squares fit to the experimental data.\label{fig:lambda_high_collapse}}
\end{center}
\end{figure}

The flat-tipped finger keeps widening as the level of collapse is increased. The relative width of the finger $\lambda$ is plotted as a function of the level of initial collapse in figure~\ref{fig:lambda_high_collapse}. $\lambda$ follows the same linear dependence on $\mathcal{A}_i/\mathcal{A}_0$ as for low levels of collapse, with the same slope. This is not surprising because the mass conservation argument exposed for lower levels of collapse still holds and the final reopening height is comparable to the one measured for lower levels of initial collapse (as can be seen from the reopening profiles in figure~\ref{fig:profiles_high_collapse}). $\mathcal{A}_i/\mathcal{A}_0$ should thus follow the same trend irrespective of the level of initial collapse, in agreement with the experimental result. \\
However, the shape at the tip undergoes smooth but significant evolution as the width of the most collapsed region increases. Closer inspection of the finger tip shows that the width of the flat interface at the finger front is equal to the width of the highly collapsed region in the centre of the cross-section (see figure~\ref{fig:tube-law} (\textit{a}) for comparison), so that the flat interface develops over the whole width of the region of very small uniform thickness. This is followed by a region where the finger broadens behind the flat leading edge, over which the channel inflates. Figure~\ref{fig:finger_high_collapse} shows that the in-plane curvature of the interface in that broadening region is very peculiar: it decreases for increased initial collapse, going from positive curvature at the POWC to almost zero, possibly even locally negative curvature for the strongest level of initial collapse achieved. Negative in-plane curvature was also clearly visible in the shape of the pointed finger reopening strongly collapsed elastic tubes at high capillary numbers (see figure~\ref{fig:pointed} (\textit{d})). \\

\begin{figure}
\begin{center}
\includegraphics[height=2.5cm]{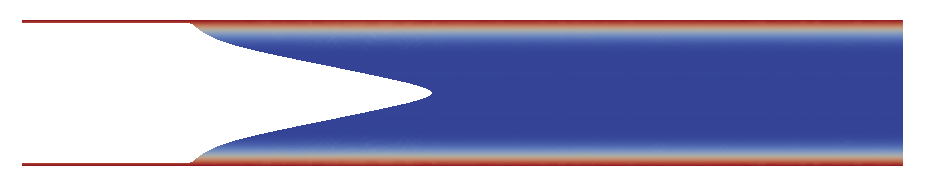}
\caption{Shape of a symmetric finger obtained by finite-element simulations (using the \texttt{oomph-lib} library) in a depth-averaged model of air finger propagation in a rigid channel of spatially varying depth. Colours map the relative local channel depth, which has been chosen to mimic the membrane profile at very high collapse. The capillary number is Ca=0.47.  \label{fig:nums}}
\end{center}
\end{figure}

Vessel reopening was conjectured to be essential to sustain the propagation of the pointed finger~\citep{heap2009bubble}. In order to assess the role of membrane reopening on the propagation of the analogous flat-tipped finger, we exploit the quasi two-dimensional nature of the flow in the elasto-rigid channel to numerically investigate the propagation of an air finger in an axially uniform rigid channel whose cross-section is similar to the profile of the highly collapsed membrane. We use the two-dimensional model for finger propagation in partially occluded rigid Hele-Shaw channels developed by~\citet{thompson2014multiple}. The model and its implementation with the finite-element library \texttt{oomph-lib} following the methods of~\citet{thompson2014multiple} are discussed in Appendix~\ref{sec:appendix}. The fluid equations are depth-averaged and both the in-plane and the cross-sectional curvature of the interface (as set by the local channel depth) are taken into account. The numerical results obtained in this framework were shown to be in quantitative agreement with the experimental data for finger propagation in partially occluded rigid channels providing that the aspect ratio of the channel was larger than 40~\citep{franco2016sensitivity}. We expect this model to give us qualitative information on the propagation modes arising when introducing a wide region of strong constriction in the middle of a rigid channel, with a similar profile to the strongly collapsed channel. \\
We mimicked the cross-section at high levels of initial collapse by taking the experimental cross-sectional profile at $\mathcal{A}/\mathcal{A}_0=0.16$ (most collapsed profile in figure~\ref{fig:tube-law} (\textit{a})) and scaling it to 85\% of its maximum height to introduce a finite small thickness for the central fluid layer. We searched for steady propagation modes and imposed symmetry with respect to the channel axis. At the experimental value of $\mathrm{Ca}=0.47$, we found a pointed finger shape illustrated in figure~\ref{fig:nums}. This propagation mode is very different from the modes previously obtained in rigid channels with smaller occlusions, as described in~\citet{franco2016sensitivity}: the in-plane curvature of this pointed shaped finger no longer decreases smoothly from a higher curvature at the finger tip to null curvature far behind it. Instead, the in-plane curvature of the finger is set far behind the tip towards the side boundaries, where the depth of the channel rapidly varies. In this region, the strong depth gradient imposes a rapid change in the cross-sectional curvature of the interface. Because of the large magnitude of the change in cross-sectional curvature, surface tension forces locally dominate over viscous effects, meaning that the total mean curvature of the interface remains approximately constant over the finger profile. As a consequence, the in-plane curvature decreases, and can even become negative, in the regions of strong negative depth gradients. In the central collapsed region, the cross-sectional curvature is constant and the in-plane curvature is zero because there is almost no flow of the fluid (the flow mostly occurs in the deeper side channels). The angle of the two long sections of straight interface is set by the in-plane curvature at the edge of the central uniformly collapsed region. The two central portions of straight interface match smoothly in the middle of the channel, where the radius of curvature at the tip scales as the local uniform channel depth.\\
We conjecture that the pointed and flat-tipped fingers are the translation of this rigid channel finger shape into, respectively, strongly collapsed compliant tube and channel geometries. In this view, the pointed finger shape is not a propagation mode that is specific to compliant vessels. Compliance, however, has several consequences on this propagation mode. Firstly, it modifies the finger shape: in compliant vessels, the curvature of the finger is also set at the edges, but imposed by the varying cross-sectional depth as the vessel is reopening, rather than the initial static profile. Besides, the tip of the pointed shaped finger observed in rigid channels cannot be sustained under compliant membranes: the pressure in the air finger initiates vessel reopening by infiltrating the thin oil layer. We would expect that elasticity would promote greater restoring capillary forces due to the taper at the finger tip, and thus that the pointed shape observed in the rigid channel would be smoothed. We observe experimentally that the tip of the finger in the elasto-rigid channel is indeed not pointed but replaced by a flat leading front whose width is set by the width of the thin liquid layer. Visualisation of the very tip of the pointed finger was not possible in the tube (the interface gets very thin, resulting in poor contrast as shown in figure~\ref{fig:pointed}), but in the light of the results obtained in the channel it is likely that the pointed finger is actually not cusp-shaped but cut to a narrow flat tip, as wide as the most collapsed part of the tube cross-section. Secondly, and more importantly, vessel compliance is necessary to stabilise this propagation mode, and make it experimentally observable: the propagation of the pointed finger in a rigid channel is unstable, as shown by the destabilisation of the computed profile when subjected to time evolution.  \\

The leading front of the flat-tipped finger is propagating into a region of high and uniform collapse, meaning that the central part of the interface experiences an effective elasto-rigid Hele-Shaw channel of uniform and very high aspect ratio. The presence of the steep lateral depth gradients at the sides of this central Hele-Shaw channel decouples the propagation of the front from the presence of the rigid channel walls, creating a convenient way of propagating a flat interface into an elasto-rigid Hele-Shaw channel. This flat interface is subject to viscous fingering on a very small scale (not visible on the pictures of figure~\ref{fig:finger_high_collapse}), well decoupled from any small in-plane curvature of the flat tip. Those small scale fingers are of constant length and similar to the ones observed in a radial elasto-rigid Hele-Shaw cell~\citep{pihler2012suppression}. The propagation of the flat-tipped finger thus allows us to study the viscous fingering of a steadily propagating flat interface in a compliant Hele-Shaw channel, as reported in~\citet{ducloue2016fingering}. \\

\subsection{Reopening from a very collapsed state: effect of $\mathrm{Ca}$}
\label{sec:Ca}
The experiments performed on the reopening of collapsed elastic tubes showed that the pointed finger could only be sustained for high levels of initial collapse and high capillary numbers. At lower capillary numbers, a variety of propagation modes were observed instead, including the asymmetric and double-tipped fingers~\citep{heap2008anomalous}. In this section, we investigate the reopening dynamics of the elasto-rigid channel from a single highly collapsed state in which the flat-tipped finger is observed at high capillary numbers, and study its evolution as the capillary number is decreased. \\

\begin{figure}
\begin{center}
\includegraphics[height=1.6cm]{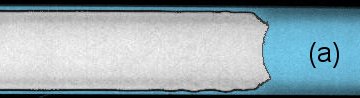}
\includegraphics[height=1.6cm]{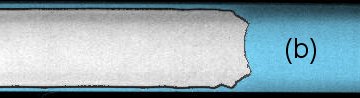}\\
\includegraphics[height=1.6cm]{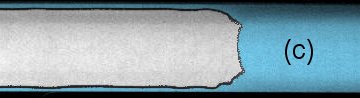}
\includegraphics[height=1.6cm]{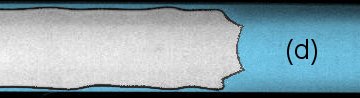}\\
\includegraphics[height=1.6cm]{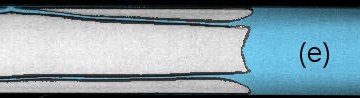}
\includegraphics[height=1.6cm]{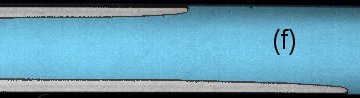}
\caption{Shape of the propagating finger for strong initial collapse ($\mathcal{A}_i/\mathcal{A}_0$=0.16) and decreasing finger speed. The corresponding values of $\mathrm{Ca}$ are: (\textit{a}) $0.51\pm0.03$, (\textit{b}) $0.35\pm0.03$, (\textit{c}) $0.21\pm0.02$, (\textit{d}) $0.11\pm0.02$, (\textit{e}) $0.09\pm0.01$ and (\textit{f}) $0.047\pm0.004$. \label{fig:finger_varCa}}
\end{center}
\end{figure}

The experimental finger shapes for the reopening of the channel at the highest level of initial collapse investigated ($\mathcal{A}_i/\mathcal{A}_0$=0.16) and across a range of capillary numbers decreasing from 0.51 to 0.047 are presented in figure~\ref{fig:finger_varCa}. The flat-tipped finger previously described is observed at high capillary numbers but disappears at low $\mathrm{Ca}$ and is replaced by two long fingers in the side channels. A similar propagation mode was observed in an elastic tube for very high levels of collapse at low capillary numbers, in which two long precursor fingers were travelling down the side lobes~\citep{heap2009bubble}. In the elasto-rigid channel, the transition from long side fingers to the flat-tipped one occurs through the growth of a central finger from the meniscus joining the two long fingers. This creates mixed states in which the central finger coexists with the side ones around a critical capillary number, as can be seen in figure~\ref{fig:finger_varCa} (\textit{e}).\\

\begin{figure}
\begin{center}
\includegraphics[height=5.5cm]{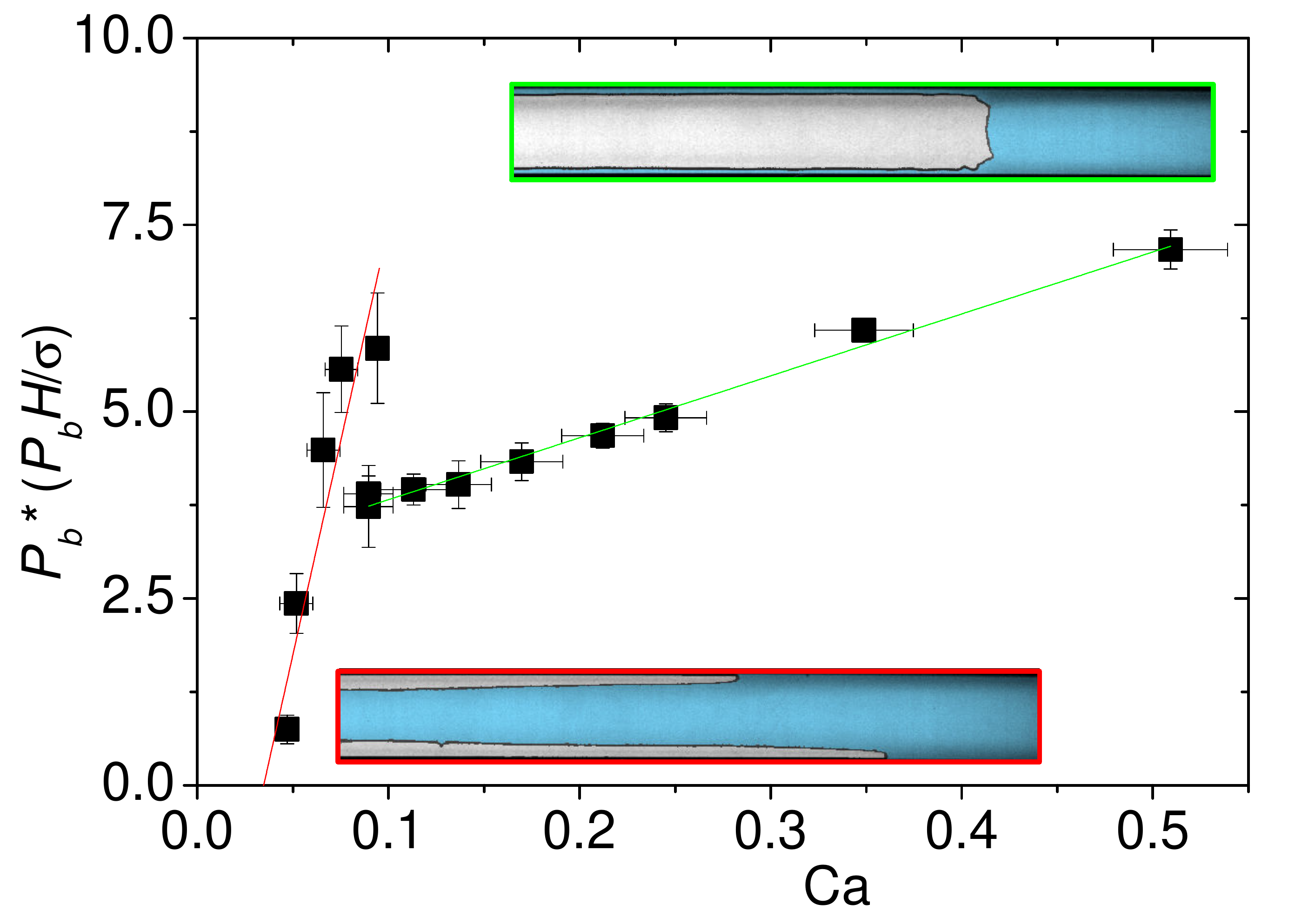}
\caption{Dimensionless air finger pressure as a function of the capillary number for reopening from a single highly collapsed initial state ($\mathcal{A}_i/\mathcal{A}_0$=0.16). Solid lines are linear fits to experimental data in each regime. Insets show representative finger shapes in each propagation mode. \label{fig:p_high_collapse}}
\end{center}
\end{figure}

The pressure of the propagating fingers is plotted as a function of the capillary number in figure~\ref{fig:p_high_collapse}, as well as a reminder of representative finger shapes obtained respectively at low and high capillary numbers. It is remarkable that the two propagation modes follow well defined $P_b-\mathrm{Ca}$ relations, with a linear trend, but that a discontinuous pressure jump (with a bi-stability region) occurs at the transition. In each propagation mode, viscous effects scale with $\mathrm{Ca}$ as shown by the linear $P_b-\mathrm{Ca}$ relations. The slope in the first regime, at low capillary numbers, is much larger than in the second regime, indicating that viscous resistance is larger in the propagation mode at low speeds. 
A very similar behaviour has been previously observed for the reopening of elastic tubes: at high levels of initial collapse, the tube is reopened by the pointed finger at high capillary numbers and the transition to the pointed finger is associated with a jump towards lower air finger pressure compared to lower capillary number reopening states (see figure~\ref{fig:pointed}). However, the elasto-rigid channel shows a simpler behaviour than the elastic tube: for capillary numbers below the transition to the flat-tipped finger, only one propagation mode is observed in the elasto-rigid channel, which minimises viscous resistance by propagating air in the deeper side channels. In the tube, on the contrary, asymmetric and double-tipped fingers followed the same smooth $P_b-\mathrm{Ca}$ curve below the critical capillary number for the transition to the pointed finger~\citep{heap2009bubble}. \\

\begin{figure}
\begin{center}
\includegraphics[height=2.5cm]{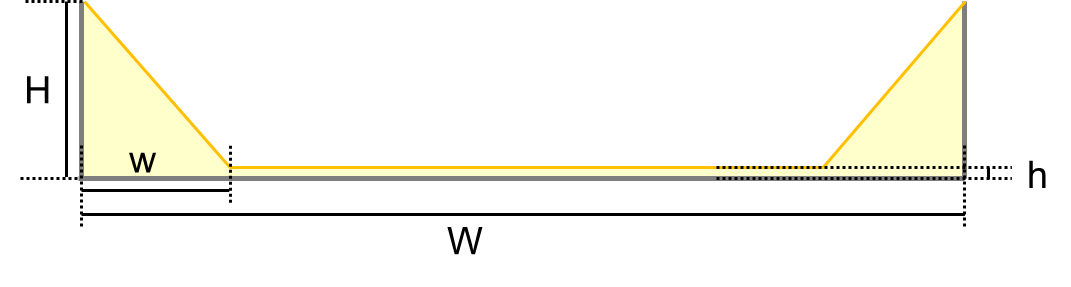}
\caption{Simplified schematic diagram of the shape taken by the membrane for very high levels of collapse. The depth $h$ of the thin liquid layer in the centre of the channel is below experimental resolution. Distances on the vertical axis have been stretched by a factor 6 for clarity.\label{fig:cross-sec} }
\end{center}
\end{figure}

Relying on the simple geometry of the elasto-rigid channel, the transition can be understood from a simple scaling argument. Pressure forces in the injected air work against the viscous dissipation in the oil and the elastic forces in the membrane as the channel inflates, which can be estimated separately for the reopening from an initially strongly collapsed state. A schematic drawing of the channel cross-section for high levels of initial collapse is presented in figure~\ref{fig:cross-sec} to define the useful lengths to characterise the membrane shape.\\

We first consider the limit case in which the channel is completely rigid, so that there is no work of elastic forces and all the work is dissipated by viscous forces in the fluid. This problem is similar to the problem of air finger propagation in rigid tubes investigated by~\citet{bretherton1961motion}. Fluid is pushed ahead of the finger tip over a characteristic length given by the length of the channel $L$. There are two limit possibilities for the finger, which can either sit in the two side channels (less viscous resistance) or propagate in the central layer (less fluid to displace).\\
For propagation in the central layer (thickness $h$, width $W-2w$), the dissipated power reads
\begin{equation}
\mathcal{P}_c\sim\mu (U/h)^2L(W-2w)h\sim\frac{\mu U^2L(W-2w)}{h}
\end{equation}
where $L(W-2w)h$ is the volume of fluid displaced and $U/h$ is the velocity gradient in the fluid. For propagation down the two side channels (width $w$, average height $H/2$), the dissipated power is
\begin{equation}
\mathcal{P}_s\sim\mu (2U/H)^2LwH\sim\frac{4\mu U^2Lw}{H}
\end{equation}
where $2U/H$ is the average velocity gradient in the triangular lobes and $LwH$ is the volume of fluid displaced. $\mathcal{P}_c/\mathcal{P}_s$ is of order $H/h$: the very thin layer in the middle of the channel leads to very high gradients compared to the sides and so we expect that for rigid channels with such geometry, the propagation will occur through two side fingers with an associated dissipated power given by $\mathcal{P}_s$. This is indeed what is seen in the experiments at very low $\mathrm{Ca}$, where little deflection of the membrane occurs.\\
We now consider the opposite limit case in which any viscous dissipation can be neglected and all the work is used for the quasi-static elastic reopening of the membrane, from $\mathcal{A/A}_0\sim0.16$ to $\mathcal{A/A}_0\sim2$. The work of the elastic forces per unit time scales as $P_{\mathrm{r}} U H W$, where $P_{\mathrm{r}}$ is the change in static pressure required to quasi-statically inflate the membrane from $\mathcal{A/A}_0\sim0.16$ to $\mathcal{A/A}_0\sim2$ and $UHW$ is the inflated volume per unit time. The channel constitutive behaviour shows that this pressure difference is $P_{\mathrm{r}}\sim1200$~Pa (see figure~\ref{fig:tube-law} (\textit{b})). \\
Balancing the power of viscous and elastic forces gives
\begin{equation}
U_{\mathrm{crit}}\sim\frac{P_{\mathrm{r}} H^2 W}{4w \mu L}
\end{equation}
or $\mathrm{Ca_{crit}}=U_{\mathrm{crit}}\mu / \sigma \sim 0.15$ using the values given in section~\ref{sec:exp} for the parameters, which gives a reasonable prediction for the experimental value of $\mathrm{Ca_{crit}}\sim 0.1$. The regime at low $\mathrm{Ca}$ is governed by minimization of viscous dissipation, whereas the one at high $\mathrm{Ca}$ is dominated by quasi-static reopening of the membrane. This is consistent with the greater importance of viscous forces at low capillary numbers pointed out when discussing the $P_b-\mathrm{Ca}$ curve. \\
The power of pressure forces in the advancing air bubble, which work again viscous dissipation in the liquid and elastic forces in the membrane, is given by $P_b U \mathcal{S}$ where $\mathcal{S}$ is the projected area of the air finger. At the transition, the power of pressure forces is the same and the speed $U$ is the same for the two reopening modes. The projected area of the flat-tipped finger, however, is similar to the one of the two long side fingers far behind the tip (the reopening heights differ by a 100 microns), but larger in the tip region, which is consistent with the pressure drop observed as soon as the flat-tipped finger appears. \\

The same transition between dominant forces occurs in an elastic tube, for which a similar scaling argument can be made for high levels of collapse using the data of~\citet{heap2009bubble}. The tube had an undeformed radius $R=5$~mm, length $\mathcal{L}=1$~m, and was filled with paraffin oil of viscosity $\eta=0.204$~$\mathrm{Pas}$ and surface tension $\Sigma=0.029$~$\mathrm{mN/m}$. The transmural pressure became very large in the highly collapsed tube, and could not be measured for the highest levels of collapse~\citep[see][figure 4]{heap2009bubble}. The largest collapse for which it is measured is $\mathcal{A/A}_0=0.12$, where $P=-1500$~$\mathrm{Pa}$. For this level of collapse, the tube takes a dumbbell shape with two side lobes of height $b=$2.5~mm~\citep{alex2008thesis}. By approximating the cross-section of the side lobes as being circular and assuming that all the flow occurs in the side lobes, the power dissipated by the viscous forces as two long fingers travel down the side lobes without deforming the tube is
\begin{equation}
\mathcal{P}^{\mathrm{tube}}_v \sim\eta \left( \frac{U}{b} \right)^2 \mathcal{L} \pi \left( \frac{b}{2} \right)^2 \times 2
\end{equation}
where we take the length of the tube as the characteristic length of flow. The tubes reopens to the neutral point~\citep{alex2008thesis}, so that the power required to quasi-statically reopen the tube in the absence of viscous effects is
\begin{equation}
\mathcal{P}^{\mathrm{tube}}_e \sim P^{\mathrm{tube}}_{\mathrm{r}} U \pi R^2
\end{equation}
with $P^{\mathrm{tube}}_{\mathrm{r}}=1500$~$\mathrm{Pa}$. Equating the two pressures leads
\begin{equation}
U_{\mathrm{crit}} \sim \frac{2P^{\mathrm{tube}}_{\mathrm{r}} R^2}{\eta \mathcal{L}}
\end{equation}
which corresponds to a critical capillary number of 2.6 for the expected transition to the pointed finger. A value of $\mathrm{Ca}\sim 2.5$ is experimentally reported~\citep[see][figure 5]{heap2009bubble}, which is consistent with our estimation. Several viscous peeling modes are observed at low capillary numbers in an elastic tube, depending on fluid distribution, and the details of the finger shape evolution from the double-tipped to the pointed finger differ from that of the long fingers to the flat-tipped one. However, this simple scaling argument shows that both transitions are governed by a change in the balance between viscous forces and elastic forces as the capillary number increases. \\

\section{Conclusions}
We have studied the steady propagation of air fingers in a collapsed elasto-rigid Hele-Shaw channel and the consequent reopening thereof. The simple constitutive behaviour of our elasto-rigid Hele-Shaw channel resulted in the experimental observation of fewer propagation modes compared to that of an elastic tube: for low levels of initial collapse, the channel was reopened by a single symmetric peeling air finger, while a flat-tipped finger was found to reopen the channel for high collapse. Precise control of the channel cross-sectional profile allowed us to show that the transition between those two finger shapes occurs smoothly as the level of initial collapse increases. The shape of the flat-tipped finger, its disappearance at low capillary numbers and the associated discontinuous pressure drop reveal that it is analogous to the pointed finger propagating at high capillary numbers in strongly collapsed elastic tubes. A simple scaling argument confirms that viscous dissipation is smaller than the work of elastic forces during the propagation of the pointed or flat-tipped fingers, whereas the finger shapes observed at lower capillary numbers in highly collapsed vessels minimise viscous resistance. Finally, numerical simulations in a rigidly collapsed channel showed that the propagation mode of the pointed or flat-tipped finger is not specific to compliant vessels, but rather triggered by steep variations in cross-sectional depth. In our system, wall elasticity stabilises this propagation mode.\\
The existence in our elasto-rigid geometry of a propagation mode analogous to the pointed finger demonstrates the potential generality of this high speed and low pressure propagation mode first uncovered in elastic tubes. It seems likely that any compliant vessel can be reopened by a similar elastically dominated propagation mode, providing that it is initially strongly collapsed. The very small amount of liquid left in the lungs at birth probably acts to select this most favourable reopening mode.\\

\begin{acknowledgments}
We thank Malcolm Walker and Carl Dixon for technical support, as well as Draga Pihler-Puzovi\'c and Matthias Heil for fruitful discussions. This work has been supported by the Leverhulme Project Grant RPG-2014-081.
\end{acknowledgments}
\newpage
\appendix
\section{Fitting curves for the self-similar membrane profiles}
\label{sec:appendixA}
\begin{figure}
\begin{center}
\includegraphics[height=9cm]{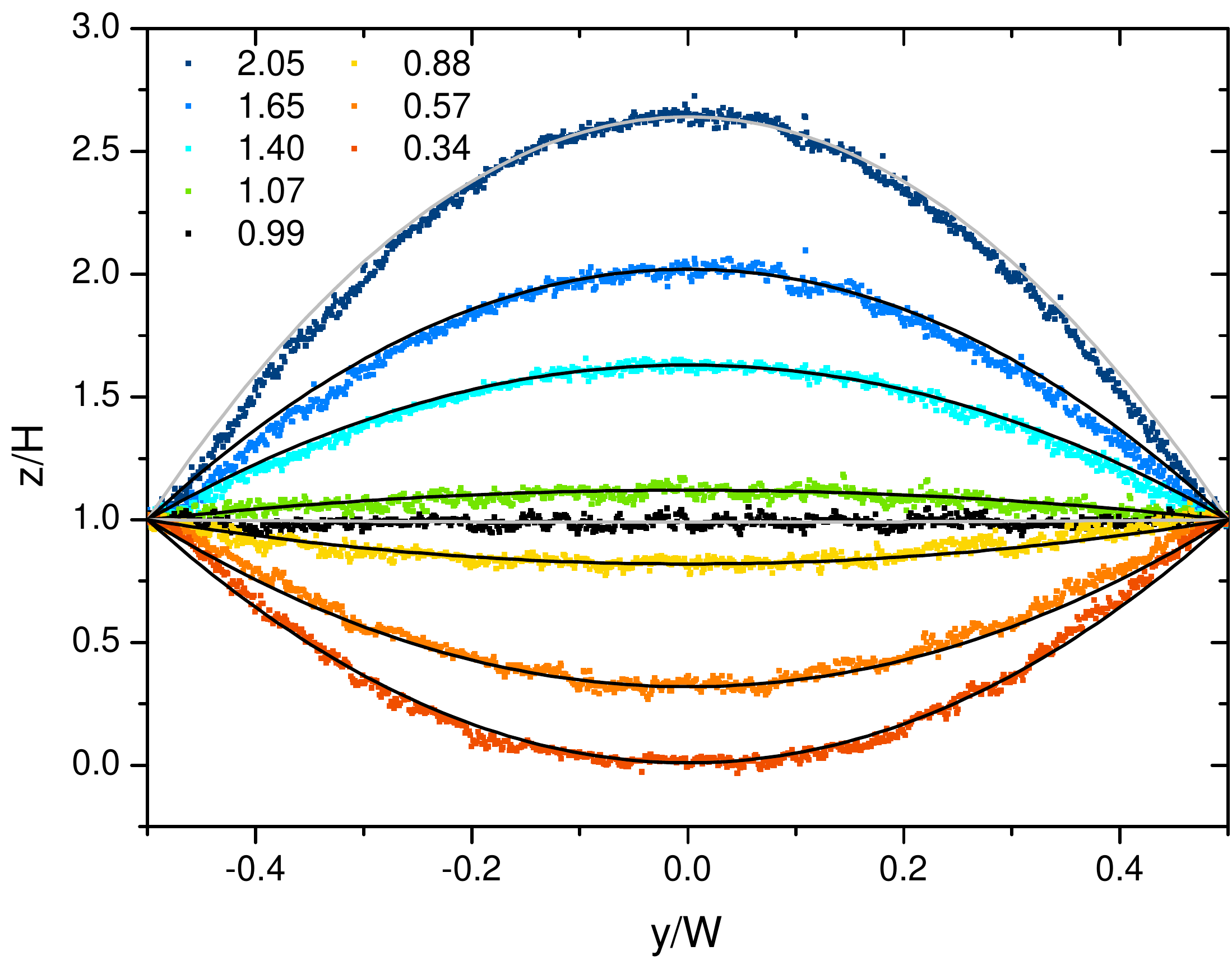}
\caption{Membrane profiles for $\mathcal{A}/\mathcal{A}_0>0.36$, overlaid with the fit functions defined in equation~\ref{eq:parabola}. The values of $\mathcal{A}/\mathcal{A}_0$ are given in the legend.\label{fig:fits}}
\end{center}
\end{figure}

An enlargement of the profiles presented in figure~\ref{fig:tube-law} (\textit{a}) is replotted for $\mathcal{A}/\mathcal{A}_0>0.36$ in figure~\ref{fig:fits}. The profiles have been fitted to polynomial functions of formula 
\begin{equation}
z/H=1-a(y/W-0.5)(y/W+0.5)
\label{eq:parabola}
\end{equation}
with $a$ a single parameter dependent on the level of collapse (overlaid solid lines). The corresponding values of $a$ are given in table~\ref{tab:a}. \\
The quality of the fit shows the similarity of the membrane profiles. A small discrepancy is noticeable on all profiles closer to the channel side walls, where bending effects become important.
\begin{table}
\begin{center}
\begin{tabular}{r c c c c c c c c}
$\mathcal{A}/\mathcal{A}_0$ & 2.05 & 1.65 & 1.40 & 1.07 & 0.99 & 0.88 & 0.57 & 0.34\\
$a$ & 6.56 & 4.08 & 2.52 & 0.48 & -0.04 & -0.72 & -2.72 & -3.96 \\
\end{tabular}
\caption{Fitting parameter $a$ as defined in equation~\ref{eq:parabola}, as a function of the level of collapse, for the profiles presented in figure~\ref{fig:fits}. \label{tab:a}}
\end{center}
\end{table}

\section{Finite-element simulations of the propagation of an air finger in a rigid constricted Hele-Shaw channel}
\label{sec:appendix}
In this appendix, we discuss briefly the model used for the computation of steady finger shapes in strongly occluded rigid channels. We refer the reader to~\citet{thompson2014multiple} for a full description of this model and its finite-element implementation. \\
\subsection{Depth-averaged model}
\begin{figure}
\begin{center}
\includegraphics[height=5cm]{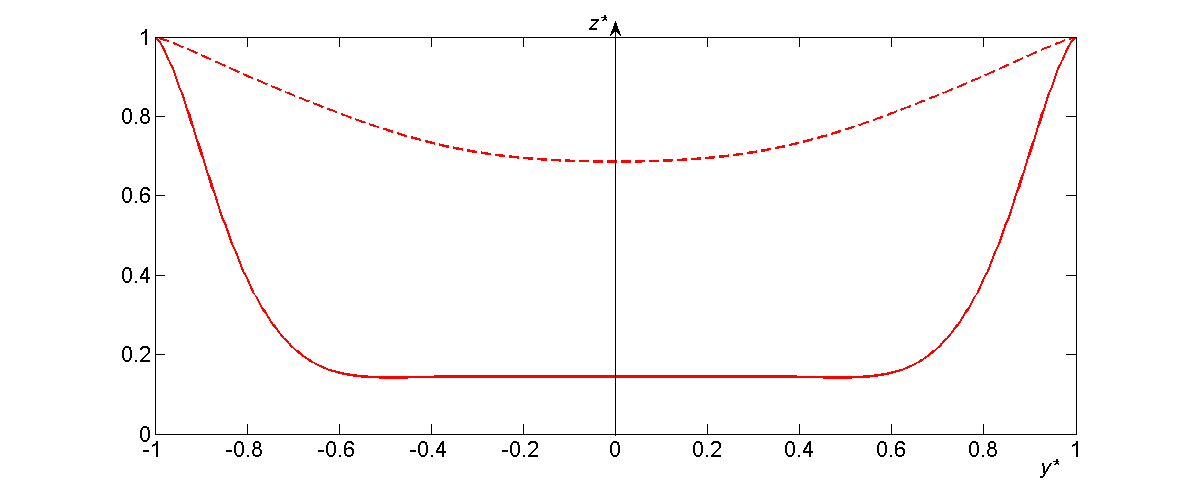}
\caption{Profiles of the cross-section used for the computations in rigid channels. Dotted line: low level of collapse; solid line: high level of collapse. The channel floor is at $z^*=0$. \label{fig:nums_Xsec}}
\end{center}
\end{figure}

We use the depth-averaged model for finger propagation in a partially occluded Hele-Shaw channel developed by~\citet{thompson2014multiple} and used by~\citet{franco2016sensitivity} who compared its results with their experimental data. We performed simulations using channel cross-sectional depth profiles derived from the experimental membrane profile at two levels of initial collapse: a high level of initial collapse ($\mathcal{A/A}_0=0.16$), used for the simulation presented in figure~\ref{fig:nums}, and a low level of initial collapse ($\mathcal{A/A}_0=0.79$) for comparison. We worked in a Cartesian system of coordinates $(x,y,z)$ in which the channel axis is aligned with $x$ and $(y,z)$ span the channel cross-section. The cross-sectional depth profile is defined by $z=b(y)$. The coordinates $x$ and $y$ were non-dimensionalised on the scale $W/2$ and the depth profile on the scale $H$. We set the aspect ratio $\alpha=W/H$ to its experimental value of 30. The experimental membrane profile was fitted to an even polynomial function of order 10 to generate a smooth and even cross-sectional profile for the simulations. This profile was used as such for the low level of collapse, and scaled to 85\% of its maximum height for the high level of collapse, in order to introduce a finite small thickness for the central liquid layer. Both profiles used for the numerical simulations are shown in dimensionless units in figure~\ref{fig:nums_Xsec}. Hereinafter, asterisks are used to distinguish dimensionless quantities from their dimensional equivalents.\\ 
The finger propagated under a pressure gradient $-G \mathbf{e}_x$ imposed far ahead of the tip. The fluid pressure $p$ was non-dimensionalised by $GW/2$ and the two components of the depth-averaged velocity $\mathbf{u}$ by $U_0=GH^2/(12\mu)$. In the lubrication approximation~\citep{reynolds1886theory}, the fluid pressure $p = p(x,y)$ is governed by
\begin{equation}
\nabla^* \cdot (b^{*3}\nabla^* p^*) = 0
\end{equation}
in the frame of reference moving with the finger tip, \textit{i.e.} with constant axial velocity $U_f$. A null pressure gradient in the $y$ direction was imposed on the channel side walls. Following~\citet{mclean1981effect}, the interface satisfies the following dimensionless boundary conditions in the $(x,y)$ plane:
\begin{equation}
p^*_b-p^*=\frac{1}{3\alpha Q} \left( \frac{1}{b^*(y^*)}+\frac{\kappa^*}{\alpha} \right)
\label{eq:laplace}
\end{equation}
\begin{equation}
\mathbf{\hat{n}}\cdot \frac{\partial \mathbf{R}^*}{\partial t^*}+U^*_f \mathbf{\hat{n}}\cdot \mathbf{e}_x+b^{*2}\mathbf{\hat{n}}\cdot\nabla^* p^* = 0
\label{eq:velocity}
\end{equation}
where $\mathbf{\hat{n}}$ is the unit vector normal to the interface and $\mathbf{R}$ is the position of a material point on the interface. Equation~\ref{eq:laplace} is the dimensionless form of the Young-Laplace equation where $Q=\mu U_0/\sigma$ is the dimensionless flow rate (equivalent to a $U_0$-based capillary number) and $\kappa^*$ is the dimensionless in-plane curvature of the finger. The contribution of viscous stresses to the pressure jump at the interface is neglected. Equation~\ref{eq:velocity} ensures the continuity of the normal component of the velocity at the interface. Symmetry was imposed by computing the finger shape over half of the channel ($y^*>0$) and imposing that the tangent to the interface was oriented along $\mathbf{e}_y$ at the finger tip. \\

\subsection{Numerical implementation}
The equations were discretised using a finite element method and the discrete residuals were assembled and solved using the finite-element library \texttt{oomph-lib}. The numerical methods were similar to those described in~\citet{thompson2014multiple}, except for the imposed symmetry. The code was validated by reproducing the shape of the finger obtained by~\citet{mclean1981effect} for $b^*=1$ as a function of the capillary number, and successfully reproducing the multiple-tipped finger shapes obtained in the full domain computations in the same range of parameters and for the same cross-sectional shapes as described by~\citet{franco2016sensitivity}. \\

To reach a solution at $\mathrm{Ca}=0.47$ with our experimental profile, we started from the static symmetric finger solution in an initially flat channel at $\alpha=30$ and $\mathrm{Ca}=0.47$. We then increased the relative constriction height step-wise, at constant capillary number, by using the pressure gradient as a free parameter. We checked that setting the pressure gradient to unity and rescaling $Q$ and $U_f$ accordingly did not modify the solution when the final desired profile had been reached.\\

\begin{figure}
\begin{center}
\includegraphics[height=2.85cm]{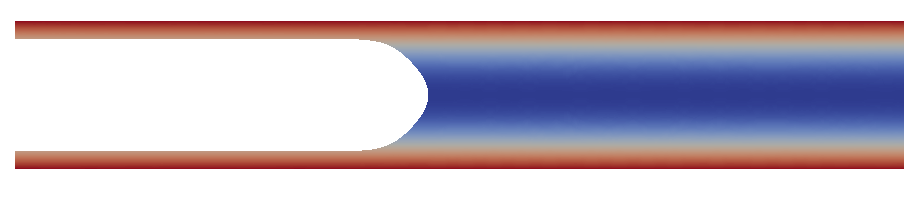}
\caption{Shape of a symmetric finger obtained by finite-element simulations (using the \texttt{oomph-lib} library) in a depth-averaged model of air finger propagation in a rigid channel of spatially varying depth. Colours map the relative local channel depth, which follows the experimental profile for $\mathcal{A/A}_0=0.79$. The capillary number is $\mathrm{Ca}=0.47$.  \label{fig:nums_app}}
\end{center}
\end{figure}

For the highly collapsed profile, we found that the finger tip became increasingly pointed as the relative height of the constriction went from 0 to 0.85. An animation showing the evolution of the finger profile at fixed $\mathrm{Ca}=0.47$ during the increase of constriction height is available as supplementary information. At a relative constriction height of 0.85, we found the finger shape presented in figure~\ref{fig:nums}.\\
We investigated a lower level of collapse to probe the sensitivity of the finger profile on the cross-sectional depth gradient. The result of the simulation for a finger propagating at $\mathrm{Ca}=0.47$ in this moderately collapsed geometry (smoothly decreasing to 70\% of the maximum height $H$ in the middle of the channel) is presented in figure~\ref{fig:nums_app}. Although the level of collapse is low, the curvature of the finger tip is most affected by the non-uniform cross-section where the depth gradient is larger.\\


%
%

%
\newpage

\bibliographystyle{jfm}
\bibliography{biblio-reopening}

\begin{thebibliography}{42}
\expandafter\ifx\csname natexlab\endcsname\relax\def\natexlab#1{#1}\fi
\def\au#1{#1} \def\ed#1{#1} \def\yr#1{#1}\def\at#1{#1}\def\jt#1{\textit{#1}}
  \def\bt#1{#1}\def\bvol#1{\textbf{#1}} \def\vol#1{#1} \def\pg#1{#1}
  \def\publ#1{#1}\def\arxiv#1{#1}\def\org#1{#1}\def\st#1{\textit{#1}}

\bibitem[Al-Housseiny \& Stone(2013)]{al2013controlling}
{\sc \au{Al-Housseiny, T.~T.} \& \au{Stone, H.~A.}} \yr{2013}  \at{Controlling
  viscous fingering in tapered {H}ele-{S}haw cells}.  \jt{Phys. Fluids}
  \bvol{25}~(9),  \pg{092102}.

\bibitem[Al-Housseiny {\em et~al.\/}(2012)Al-Housseiny, Tsai \&
  Stone]{al2012control}
{\sc \au{Al-Housseiny, T.~T.}, \au{Tsai, P.~A.} \& \au{Stone, H.~A.}} \yr{2012}
   \at{Control of interfacial instabilities using flow geometry}.  \jt{Nat.
  Phys.}  \bvol{8}~(10),  \pg{747--750}.

\bibitem[Ben~Amar {\em et~al.\/}(1993)Ben~Amar, Combescot \&
  Couder]{amar1993viscous}
{\sc \au{Ben~Amar, M.}, \au{Combescot, R.} \& \au{Couder, Y.}} \yr{1993}
  \at{Viscous fingering with adverse anisotropy: A new {S}affman-{T}aylor
  finger}.  \jt{Phys. Rev. Lett.}  \bvol{70}~(20),  \pg{3047}.

\bibitem[Bensimon(1986)]{bensimon1986stability}
{\sc \au{Bensimon, D.}} \yr{1986}  \at{Stability of viscous fingering}.
  \jt{Phys. Rev. A}  \bvol{33}~(2),  \pg{1302}.

\bibitem[Bland(2001)]{bland2001loss}
{\sc \au{Bland, R.~D.}} \yr{2001}  \at{Loss of liquid from the lung lumen in
  labor: more than a simple “squeeze”}.  \jt{Am. J. Physiol.-Lung C.}
  \bvol{280}~(4),  \pg{L602--L605}.

\bibitem[Borno {\em et~al.\/}(2006)Borno, Steinmeyer \&
  Maharbiz]{borno2006transpiration}
{\sc \au{Borno, R.~T.}, \au{Steinmeyer, J.~D.} \& \au{Maharbiz, M.~M.}}
  \yr{2006}  \at{Transpiration actuation: the design, fabrication and
  characterization of biomimetic microactuators driven by the surface tension
  of water}.  \jt{J. Micromech. Microeng.}  \bvol{16}~(11),  \pg{2375}.

\bibitem[Bretherton(1961)]{bretherton1961motion}
{\sc \au{Bretherton, F.~P.}} \yr{1961}  \at{The motion of long bubbles in
  tubes}.  \jt{J. Fluid Mech.}  \bvol{10}~(2),  \pg{166--188}.

\bibitem[Burger \& Macklem(1968)]{burger1968airway}
{\sc \au{Burger, E.~J.} \& \au{Macklem, P.}} \yr{1968}  \at{Airway closure:
  demonstration by breathing 100 percent {0}$_2$ at low lung volumes and by
  {N}$_2$ washout.}  \jt{J. Appl. Physiol.}  \bvol{25}~(2),  \pg{139--148}.

\bibitem[Combescot(1994)]{combescot1994saffman}
{\sc \au{Combescot, R.}} \yr{1994}  \at{Saffman-{T}aylor fingers with adverse
  anisotropic surface tension}.  \jt{Phys. Rev. E}  \bvol{49}~(5),  \pg{4172}.

\bibitem[Couder {\em et~al.\/}(1986)Couder, Gerard \& Rabaud]{couder1986narrow}
{\sc \au{Couder, Y.}, \au{Gerard, N.} \& \au{Rabaud, M.}} \yr{1986}  \at{Narrow
  fingers in the {S}affman-{T}aylor instability}.  \jt{Phys. Rev. A}
  \bvol{34}~(6),  \pg{5175}.

\bibitem[Duclou\'e {\em et~al.\/}(2016)Duclou\'e, Hazel \&
  Juel]{ducloue2016fingering}
{\sc \au{Duclou\'e, L.}, \au{Hazel, A.~L.} \& \au{Juel, A.}} \yr{2016}
  \at{Fingering and dendritic growth in a compliant channel}.  \jt{in
  preparation} .

\bibitem[Flaherty {\em et~al.\/}(1972)Flaherty, Keller \&
  Rubinow]{flaherty1972post}
{\sc \au{Flaherty, J.~E.}, \au{Keller, J.~B.} \& \au{Rubinow, S.~I.}} \yr{1972}
   \at{Post buckling behavior of elastic tubes and rings with opposite sides in
  contact}.  \jt{SIAM J. Appl. Math.}  \bvol{23}~(4),  \pg{446--455}.

\bibitem[Franco-G\'omez {\em et~al.\/}(2016)Franco-G\'omez, Thompson, Hazel \&
  Juel]{franco2016sensitivity}
{\sc \au{Franco-G\'omez, A.}, \au{Thompson, A.~B.}, \au{Hazel, A.~L.} \&
  \au{Juel, A.}} \yr{2016}  \at{Sensitivity of {S}affman-{T}aylor fingers to
  channel depth perturbations}.  \jt{J. Fluid Mech.}  \bvol{794},
  \pg{343--368}.

\bibitem[Gaver~III {\em et~al.\/}(1996)Gaver~III, Halpern, Jensen \&
  Grotberg]{gaver1996steady}
{\sc \au{Gaver~III, D.~P.}, \au{Halpern, D.}, \au{Jensen, O.~E.} \&
  \au{Grotberg, J.~B.}} \yr{1996}  \at{The steady motion of a semi-infinite
  bubble through a flexible-walled channel}.  \jt{J. Fluid Mech.}  \bvol{319},
  \pg{25--65}.

\bibitem[Gaver~III {\em et~al.\/}(1990)Gaver~III, Samsel \&
  Solway]{gaver1990effects}
{\sc \au{Gaver~III, D.~P.}, \au{Samsel, R.~W.} \& \au{Solway, J.}} \yr{1990}
  \at{Effects of surface tension and viscosity on airway reopening}.  \jt{J.
  Appl. Physiol.}  \bvol{69}~(1),  \pg{74--85}.

\bibitem[Grotberg \& Jensen(2004)]{grotberg2004biofluid}
{\sc \au{Grotberg, J.~B.} \& \au{Jensen, O.~E.}} \yr{2004}  \at{Biofluid
  mechanics in flexible tubes}.  \jt{Ann. Rev. Fluid Mech.}  \bvol{36}~(1),
  \pg{121}.

\bibitem[Halpern \& Grotberg(1992)]{halpern1992fluid}
{\sc \au{Halpern, D.} \& \au{Grotberg, J.~B.}} \yr{1992}  \at{Fluid-elastic
  instabilities of liquid-lined flexible tubes}.  \jt{J. Fluid Mech.}
  \bvol{244},  \pg{615--632}.

\bibitem[Hazel \& Heil(2003)]{hazel2003three}
{\sc \au{Hazel, A.~L.} \& \au{Heil, M.}} \yr{2003}  \at{Three-dimensional
  airway reopening: the steady propagation of a semi-infinite bubble into a
  buckled elastic tube}.  \jt{J. Fluid Mech.}  \bvol{478},  \pg{47--70}.

\bibitem[Heap(2008)]{alex2008thesis}
{\sc \au{Heap, A.}} \yr{2008}  \at{The reopening of a collapsed, fluid-filled
  elastic tube}. PhD thesis, University of Manchester.

\bibitem[Heap \& Juel(2008)]{heap2008anomalous}
{\sc \au{Heap, A.} \& \au{Juel, A.}} \yr{2008}  \at{Anomalous bubble
  propagation in elastic tubes}.  \jt{Phys. Fluids}  \bvol{20}~(8),
  \pg{081702}.

\bibitem[Heap \& Juel(2009)]{heap2009bubble}
{\sc \au{Heap, A.} \& \au{Juel, A.}} \yr{2009}  \at{Bubble transitions in
  strongly collapsed elastic tubes}.  \jt{J. Fluid Mech.}  \bvol{633},
  \pg{485--507}.

\bibitem[Heil(1999)]{heil1999minimal}
{\sc \au{Heil, M.}} \yr{1999}  \at{Minimal liquid bridges in
  non-axisymmetrically buckled elastic tubes}.  \jt{J. Fluid Mech.}
  \bvol{380},  \pg{309--337}.

\bibitem[Hoberg {\em et~al.\/}(2014)Hoberg, Verneuil \&
  Hosoi]{hoberg2014elastocapillary}
{\sc \au{Hoberg, T.~B.}, \au{Verneuil, E.} \& \au{Hosoi, A.~E.}} \yr{2014}
  \at{Elastocapillary flows in flexible tubes}.  \jt{Phys. Fluids}
  \bvol{26}~(12),  \pg{122103}.

\bibitem[Hodson(1991)]{hodson1991first}
{\sc \au{Hodson, W.~A.}} \yr{1991} {\em The first breath\/},  \pg{pp.
  1665--1675}.  \publ{Raven New York}.

\bibitem[Jensen {\em et~al.\/}(1987)Jensen, Libchaber, Pelc{\'e} \&
  Zocchi]{jensen1987effect}
{\sc \au{Jensen, M.~H.}, \au{Libchaber, A.}, \au{Pelc{\'e}, P.} \& \au{Zocchi,
  G.}} \yr{1987}  \at{Effect of gravity on the {S}affman-{T}aylor meniscus:
  {T}heory and experiment}.  \jt{Phys. Rev. A}  \bvol{35}~(5),  \pg{2221}.

\bibitem[Jensen {\em et~al.\/}(2002)Jensen, Horsburgh, Halpern \&
  Gaver~III]{jensen2002steady}
{\sc \au{Jensen, O.~E.}, \au{Horsburgh, M.~K.}, \au{Halpern, D.} \&
  \au{Gaver~III, D.~P.}} \yr{2002}  \at{The steady propagation of a bubble in a
  flexible-walled channel: asymptotic and computational models}.  \jt{Phys.
  Fluids}  \bvol{14}~(2),  \pg{443--457}.

\bibitem[Juel \& Heap(2007)]{juel2007reopening}
{\sc \au{Juel, A.} \& \au{Heap, A.}} \yr{2007}  \at{The reopening of a
  collapsed fluid-filled elastic tube}.  \jt{J. Fluid Mech.}  \bvol{572},
  \pg{287--310}.

\bibitem[de~L{\'o}zar {\em et~al.\/}(2009)de~L{\'o}zar, Heap, Box, Hazel \&
  Juel]{de2009tube}
{\sc \au{de~L{\'o}zar, A.}, \au{Heap, A.}, \au{Box, F.}, \au{Hazel, A.~L.} \&
  \au{Juel, A.}} \yr{2009}  \at{Tube geometry can force switchlike transitions
  in the behavior of propagating bubbles}.  \jt{Phys. Fluids}  \bvol{21}~(10),
  \pg{101702}.

\bibitem[Macklem {\em et~al.\/}(1970)Macklem, Proctor \&
  Hogg]{macklem1970stability}
{\sc \au{Macklem, P.~T.}, \au{Proctor, D.~F.} \& \au{Hogg, J.~C.}} \yr{1970}
  \at{The stability of peripheral airways}.  \jt{Resp. Physiol.}  \bvol{8}~(2),
   \pg{191--203}.

\bibitem[McLean \& Saffman(1981)]{mclean1981effect}
{\sc \au{McLean, J.~W.} \& \au{Saffman, P.~G.}} \yr{1981}  \at{The effect of
  surface tension on the shape of fingers in a {H}ele {S}haw cell}.  \jt{J.
  Fluid Mech.}  \bvol{102},  \pg{455--469}.

\bibitem[Park \& Homsy(1984)]{park1984two}
{\sc \au{Park, C.-W.} \& \au{Homsy, G.~M.}} \yr{1984}  \at{Two-phase
  displacement in {H}ele {S}haw cells: theory}.  \jt{J. Fluid Mech.}
  \bvol{139},  \pg{291--308}.

\bibitem[Pihler-Puzovi{\'c} {\em et~al.\/}(2012)Pihler-Puzovi{\'c}, Illien,
  Heil \& Juel]{pihler2012suppression}
{\sc \au{Pihler-Puzovi{\'c}, D.}, \au{Illien, P.}, \au{Heil, M.} \& \au{Juel,
  A.}} \yr{2012}  \at{Suppression of complex fingerlike patterns at the
  interface between air and a viscous fluid by elastic membranes}.  \jt{Phys.
  Rev. Lett.}  \bvol{108}~(7),  \pg{074502}.

\bibitem[Pihler-Puzovi{\'c} {\em et~al.\/}(2015)Pihler-Puzovi{\'c}, Juel, Peng,
  Lister \& Heil]{pihler2015displacement}
{\sc \au{Pihler-Puzovi{\'c}, D.}, \au{Juel, A.}, \au{Peng, G.~G.}, \au{Lister,
  J.~R.} \& \au{Heil, M.}} \yr{2015}  \at{Displacement flows under elastic
  membranes. {P}art 1. {E}xperiments and direct numerical simulations}.  \jt{J.
  Fluid Mech.}  \bvol{784},  \pg{487--511}.

\bibitem[Pihler-Puzovi\'c {\em et~al.\/}(2013)Pihler-Puzovi\'c, P{\'e}rillat,
  Russell, Juel \& Heil]{pihler2013modelling}
{\sc \au{Pihler-Puzovi\'c, D.}, \au{P{\'e}rillat, R.}, \au{Russell, M.},
  \au{Juel, A.} \& \au{Heil, M.}} \yr{2013}  \at{Modelling the suppression of
  viscous fingering in elastic-walled {H}ele-{S}haw cells}.  \jt{J. Fluid
  Mech.}  \bvol{731},  \pg{162--183}.

\bibitem[Pokroy {\em et~al.\/}(2009)Pokroy, Kang, Mahadevan \&
  Aizenberg]{pokroy2009self}
{\sc \au{Pokroy, B.}, \au{Kang, S.~H.}, \au{Mahadevan, L.} \& \au{Aizenberg,
  J.}} \yr{2009}  \at{Self-organization of a mesoscale bristle into ordered,
  hierarchical helical assemblies}.  \jt{Science}  \bvol{323}~(5911),
  \pg{237--240}.

\bibitem[Reynolds(1886)]{reynolds1886theory}
{\sc \au{Reynolds, O.}} \yr{1886}  \at{On the theory of lubrication and its
  application to {M}r. {B}eauchamp tower's experiments, including an
  experimental determination of the viscosity of olive oil.}  \jt{P. Roy. Soc.
  Lond. A Mat.}  \bvol{40}~(242-245),  \pg{191--203}.

\bibitem[Roman \& Bico(2010)]{roman2010elasto}
{\sc \au{Roman, B.} \& \au{Bico, J.}} \yr{2010}  \at{Elasto-capillarity:
  deforming an elastic structure with a liquid droplet}.  \jt{J. Phys.:
  Condens. Mat.}  \bvol{22}~(49),  \pg{493101}.

\bibitem[Saffman \& Taylor(1958)]{saffman1958penetration}
{\sc \au{Saffman, P.~G.} \& \au{Taylor, G.}} \yr{1958} The penetration of a
  fluid into a porous medium or {H}ele-{S}haw cell containing a more viscous
  liquid.  \bt{In {\em P. Roy. Soc. Lond. A: Mat.\/}}, ,  \vol{vol. 245},
  \pg{pp. 312--329}. The Royal Society.

\bibitem[Shapiro(1977)]{shapiro1977steady}
{\sc \au{Shapiro, A.~H.}} \yr{1977}  \at{Steady flow in collapsible tubes}.
  \jt{J. Biomech. Eng.}  \bvol{99}~(3),  \pg{126--147}.

\bibitem[Tabeling {\em et~al.\/}(1987)Tabeling, Zocchi \&
  Libchaber]{tabeling1987experimental}
{\sc \au{Tabeling, P.}, \au{Zocchi, G.} \& \au{Libchaber, A.}} \yr{1987}
  \at{An experimental study of the {S}affman-{T}aylor instability}.  \jt{J.
  Fluid Mech.}  \bvol{177},  \pg{67--82}.

\bibitem[Thompson {\em et~al.\/}(2014)Thompson, Juel \&
  Hazel]{thompson2014multiple}
{\sc \au{Thompson, A.~B.}, \au{Juel, A.} \& \au{Hazel, A.~L.}} \yr{2014}
  \at{Multiple finger propagation modes in {H}ele-{S}haw channels of variable
  depth}.  \jt{J. Fluid Mech.}  \bvol{746},  \pg{123--164}.

\bibitem[Zhao {\em et~al.\/}(1992)Zhao, Casademunt, Yeung \&
  Maher]{zhao1992perturbing}
{\sc \au{Zhao, H.}, \au{Casademunt, J.}, \au{Yeung, C.} \& \au{Maher, J.~V.}}
  \yr{1992}  \at{Perturbing {H}ele-{S}haw flow with a small gap gradient}.
  \jt{Phys. Rev. A}  \bvol{45},  \pg{2455--2460}.

\end{thebibliography}

\end{document}